\documentclass[11pt,twoside]{article}
\pdfoutput=1

\usepackage{pslatex}    % fit more on page using PS-fonts 
\usepackage{amsfonts}
\usepackage{amsmath}
\usepackage{amsthm}
\usepackage{amscd}
\usepackage{cite}
\usepackage{eucal}
\usepackage{graphicx}
\usepackage{a4}
\usepackage{booktabs}
\usepackage{microtype}
\def\beq{\begin{equation}}
\def\eeq{\end{equation}}
\def\bea{\begin{eqnarray}}
\def\eea{\end{eqnarray}}

\def\beann{\begin{eqnarray*}}
\def\eeann{\end{eqnarray*}}

\let\a=\alpha \let\be=\beta  \let\de=\delta
\let\e=\varepsilon \let\z=\zeta \let\h=\eta 
\let\eps=\epsilon
  \let\la=\lambda \let\m=\mu
\let\n=\nu \let\x=\xi \let\p=\pi \let\r=\rho \let\s=\sigma
\let\om=\omega 
 \let\Ph=\phi  \let\Ps=\Psi
  \let\Th=\Theta
\let\La=\Lambda \let\G=\Gamma \let\D=\Delta

\let\qd=\quad \let\qqd=\qquad 

\def\epp{\, .}
\def\epc{\, ,}

\def\tst#1{{\textstyle #1}}

\theoremstyle{plain}

\newtheorem*{corollary*}{Corollary}

\newtheorem*{conjecture*}{Conjecture}

\theoremstyle{definition}

\def\2{\frac{1}{2}} \def\4{\frac{1}{4}}

\def\6{\partial}

\def\+{\dagger}

\def\<{\langle} \def\>{\rangle}

\let\then=\Rightarrow

\def\CO{{\cal O}} 
\def\CR{{\cal R}} \def\CS{{\cal S}}

\def\CC{{\cal C}} \def\CCq{\overline{\cal C}}

\def\i{{\rm i}}

\def\rd{{\rm d}}
\def\re{{\rm e}}

\DeclareMathOperator{\ch}{ch}

\DeclareMathOperator{\tr}{tr}

\DeclareMathOperator{\sign}{sign}
\DeclareMathOperator{\End}{End}

\DeclareMathOperator{\res}{res}
\DeclareMathOperator{\card}{card}

\DeclareMathOperator{\detq}{{\rm det}_q}

 \def\Im{{\rm Im\,}}

\def\xv{\mathbf{x}}

\def\Af{\mathfrak{A}}
\def\Afq{\overline{\mathfrak{A}}}
\def\af{\mathfrak{a}}

\def\fa{\mathfrak{a}}
\def\faq{\overline{\mathfrak{a}}}
\def\Bf{\mathfrak{B}}
\def\Bfq{\overline{\mathfrak{B}}}
\def\fb{\mathfrak{b}}
\def\fbq{\overline{\mathfrak{b}}}
\def\ff{\mathfrak{f}}
\def\ffq{\overline{\mathfrak{f}}}
\def\Ff{\mathfrak{F}}
\def\Ffq{\overline{\mathfrak{F}}}

\def\afII {\mathfrak{a}_{\rm II}}
\def\afIIq{\overline{\mathfrak{a}}_{\rm II}}
\def\jg2{K}
\def\kmat{\widehat {\cal K}}

\newcommand{\F}{F^{[1]}}
\newcommand{\Fq}{\overline{F}^{[1]}}

\renewcommand{\appendix}{%
   \renewcommand{\section}{%\newpage%
        \secdef\Appendix\sAppendix}%
   \setcounter{section}{0}%
   \renewcommand{\thesection}{\Alph{section}}%
   \renewcommand{\theequation}{\thesection.\arabic{equation}}%
}

\newcommand{\Appendix}[2][?]{%
     \refstepcounter{section}%
     \setcounter{equation}{0}%
     \addcontentsline{toc}{appendix}%
          {\protect\numberline{\appendixname~\thesection} #1}%
     \vspace{\baselineskip}%
     {\noindent\large\bfseries\appendixname\ \thesection: #2\par}%
     \sectionmark{#1}\vspace{\baselineskip}}

\newcommand{\sAppendix}[1]{%
     {\noindent\large\bfseries\appendixname\:: #1\par}%
     \sectionmark{#1}\vspace{\baselineskip}}

\renewcommand{\arraystretch}{1.1}

\begin{document}

\thispagestyle{empty}

\begin{center}

{\Large {\bf Correlation functions of the integrable isotropic
spin-1 chain at finite temperature
\\}}

\vspace{7mm}

{\large
Frank G\"{o}hmann\footnote[2]{e-mail: goehmann@physik.uni-wuppertal.de}}%
\\[1ex]
Fachbereich C -- Physik, Bergische Universit\"at Wuppertal,\\
42097 Wuppertal, Germany\\[2.5ex]
{\large Alexander Seel\footnote{e-mail:
alexander.seel@itp.uni-hannover.de}}\\[.5ex]
Institut f\"ur Theoretische Physik, Universit\"at Hannover,\\
Appelstr.\ 2, 30167 Hannover, Germany\\[2.5ex]
{\large Junji Suzuki\footnote[3]{e-mail: sjsuzuk@ipc.shizuoka.ac.jp}}%
\\[.5ex]
Department of Physics, Faculty of Science, Shizuoka University,\\
Ohya 836, Suruga, Shizuoka, Japan

\vspace{40mm}

{\large {\bf Abstract}}

\end{center}

\begin{list}{}{\addtolength{\rightmargin}{9mm}
               \addtolength{\topsep}{-5mm}}
\item
We represent the density matrix of a finite segment of the integrable
iso\-tropic spin-1 chain in the thermodynamic limit as a multiple
integral. Our integral formula is valid at finite temperature and also
includes a homogeneous magnetic field.
\\[2ex]
{\it PACS: 05.30.-d, 75.10.Pq}
\end{list}

\clearpage

\section{Introduction}
In recent years we have witnessed rapid progress in the understanding
of the mathematical structure of the static correlation functions
of Yang-Baxter integrable quantum systems. Most of this progress
was obtained with the example of the XXZ spin-$\2$ chain. For the
XXZ chain a hidden Grassmann structure was identified in \cite{BJMST06b,
BJMST08a} which made it possible to prove the complete factorization
of the correlation functions under very general conditions \cite{JMS08},
including the case of finite temperature and magnetic field \cite{JMS08,%
BoGo09}.

At the outset of this new development were multiple integral
representations for the density matrix of a finite chain segment
\cite{JMMN92,JiMi96,KMT99b,GKS05} and the observation in \cite{BoKo01}
that these integrals factorize into sums over products of single
integrals. With \cite{BGKS06,DGHK07} it became apparent that the
factorization is not a property of the ground state in the thermodynamic
limit, but can be done for finite temperature and for the ground states
of finite chains as well. This was part of the motivation for the
research leading to \cite{BJMST06b,BJMST08a,JMS08,BoGo09}. The multiple
integral representations also served as the starting point for a direct
calculation of the asymptotics of the ground state correlation functions
of the XXZ chain in \cite{KKMST09}.

At the present stage of research it is an interesting question to which
extend the results for the XXZ spin-$\2$ chain can be generalized to
other integrable models. The models closest to the spin-$\2$ XXZ chain
are those with the same $R$-matrix, notably the Bose gas and the
Sine-Gordon model. For both of these, partial results could be obtained
\cite{KKMST07,SBGK07a,SGK07,JMS09pp} in certain scaling limits. Another
class of models, which is closely related to the spin-$\2$ XXZ chain as
well, is the class of its higher-spin generalizations constructed by means
of the fusion procedure \cite{KuSk82b,KRS81}.

For the fused spin chains N.~Kitanine constructed a multiple integral
representation \cite{Kitanine01} for the ground state correlation
functions. He observed that much of the necessary algebraic and
combinatorial work can be carried over rather directly from the spin-$\2$
case \cite{KMT99b}. But due to the different structure of the ground state,
which is build up of strings of Bethe roots for the higher spin integrable
chains, the rewriting of the combinatorial sums as integrals in the
thermodynamic limit required some modification as compared to the spin-$\2$
case. As a result the number of integrals in Kitanine's formula is $2ms$
for the $m$-site density matrix of the spin-$s$ chain, and a subtle
regularization determines the relative location of the integration
contours. Unlike in the spin-$\2$ case his multiple integral formula
for higher spins bears no obvious similarity with the formulae obtained
within the $q$-vertex operator approach \cite{Idzumi94,BoWe94}. For
simplicity Kitanine concentrated on the isotropic (or XXX-) case, and he
did not include a magnetic field. The generalization of his work to the
XXZ-case (without magnetic field) was recently obtained in \cite{DeMa10}.

It is the aim of this work to extend Kitanine's result, exemplarily
in the simplest case of the isotropic spin-1 chain, to finite
temperatures. We shall also include a magnetic field into the calculation.
Again fusion allows us to start with spin-$\2$ and to use the algebraic
and combinatorial results of \cite{KMT99b,GHS05}. Then, as we shall see,
the crucial problem is the analytic part of the calculation, where
the combinatorial sums are converted into a multiple integral over certain
contours by means of appropriate functions.

A priori it is unclear how to choose these functions. They should be
related to the functions appearing in the description of the
thermodynamics of the spin chains. Yet, there are several mathematically
rather different formulations of the thermodynamics using different types
of auxiliary functions. In the study of the spin-$\2$ chain \cite{GKS04a,%
GHS05} only one of these formulations turned out to be compatible with
the multiple integral representation. It is the formulation based
on the quantum transfer matrix \cite{Suzuki85} and using only a finite
number of auxiliary functions which satisfy a closed set of functional
equations \cite{Kluemper93}. So far this is the least canonical
formulation. No general scheme for it is known. Fortunately, the best
understood case is just the case of the higher-spin XXX chains, which
was worked out by one of the authors \cite{Suzuki99}. As we shall see
below the auxiliary functions introduced in \cite{Suzuki99} are indeed
most useful also in the framework of multiple integral representations.
These functions can be efficiently calculated from a set of nonlinear
coupled integral equations and allow for an accurate numerical description
of the thermodynamics of the higher-spin chains \cite{Suzuki99}.
Besides the auxiliary functions that satisfy nonlinear integral equations
we shall introduce new functions, solving linear integral equations,
which will finally allow us to rewrite the combinatorial sums representing
the density matrix as a single multiple integral.

We see this work as a feasibility study and therefore stick with the
simplest higher-spin generalization of a finite-temperature multiple
integral representation. Further generalizations to general higher
spin, to the XXZ case or to include a disorder para\-meter into the
calculation are left for future studies.

The paper is organized as follows. In section 2 we recall the construction
of the Hamiltonian and the statistical operator by means of fusion
of spin-$\2$ transfer matrices. We also recall how to calculate
the density matrix of a chain segment within the quantum transfer matrix
approach. In section 3 we review the calculation of the thermodynamic
quantities by means of nonlinear integral equations and present
an alternative closed contour form of such equations. Section 4
contains our main result, which is a multiple integral formula for the
inhomogeneous density matrix of a finite chain segment. In section 5
we present a factorized form of our formulae for the one-point functions.
Finally, the zero temperature limit is sketched in section 6.
The technical details of the derivation of the nonlinear integral
equations and of the multiple integral formula have been separated from the
main text and are summarized in three appendices.

\section{Hamiltonian and density matrix}
\subsection{Hamiltonian}
The Hamiltonian of the integrable isotropic spin-1 chain on a lattice
of $2L$ sites is
\begin{equation} \label{ham}
     H = \frac{J}{4} \sum_{n = - L + 1}^L
           \bigl( S_{n-1}^\a S_n^\a - (S_{n-1}^\a S_n^\a)^2
                  \bigr) \epp
\end{equation}
Here implicit summation over $\a = x, y, z$ is understood, and periodic
boundary conditions, $S_{-L}^\a = S_L^\a$, are employed for the explicit
sum over $n$. The $S_n^\a$ act locally as standard spin-1 operators, and
antiferromagnetic exchange, $J > 0$, is assumed throughout the paper.

The Hamiltonian (\ref{ham}) was first obtained in a more general
anisotropic form in \cite{ZaFa80}. Shortly later it was constructed
by means of the fusion procedure \cite{KuSk82b,KRS81}. The ground state
and the elementary excitations were studied in \cite{Takhtajan82},
and an algebraic Bethe ansatz and the thermodynamics within the TBA
approach were obtained in \cite{Babujian82}.

\subsection{Integrable structure}
The model can be constructed by means of the fusion procedure
\cite{KRS81}, starting from the fundamental spin-$\2$ $R$-matrix
\begin{equation}
     R^{[1,1]} (\la) = \begin{pmatrix} 1 &&& \\ & b(\la) & c(\la) & \\
                              & c(\la) & b(\la) & \\ &&& 1
              \end{pmatrix} \epc \qd
	      b(\la) = \frac{\la}{\la + 2 \i} \epc \qd
	      c(\la) = \frac{2 \i}{\la + 2 \i} \epp
\end{equation}
which we think of as an element of $\End ({\mathbb C}^2 \otimes
{\mathbb C}^2)$. It satisfies the Yang-Baxter equation
\begin{equation} \label{ybe}
     R^{[1,1]}_{12} (\la - \m) R^{[1,1]}_{13} (\la) R^{[1,1]}_{23} (\m)
      = R^{[1,1]}_{23} (\m) R^{[1,1]}_{13} (\la) R^{[1,1]}_{12} (\la - \m)
        \epp
\end{equation}
As usual the $R^{[1,1]}_{jk}$ in this equation act on the $j$th and $k$th
factor of the triple tensor product ${\mathbb C}^2 \otimes {\mathbb C}^2
\otimes {\mathbb C}^2$ as $R^{[1,1]}$ and on the remaining factor
trivially. $R^{[1,1]}$ is normalized in such a way that
\begin{equation}
     R^{[1,1]} (0) = P^{[1]} \epc
\end{equation}
where $P^{[1]}$ is the transposition of the two factors in ${\mathbb C}^2
\otimes {\mathbb C}^2$. We say that $R^{[1,1]}$ is regular. At the same
time $\check R^{[1,1]} = P^{[1]} R^{[1,1]}$ satisfies the unitarity
condition
\begin{equation} \label{uni1}
     \check R^{[1,1]} (\la) \, \check R^{[1,1]} (-\la) = I_4 \epc
\end{equation} 
with $I_n$ denoting the $n \times n$ unit matrix.

A further property of $R^{[1,1]}$, which is at the heart of the fusion
procedure, is its degeneracy at two special points,
\begin{equation} \label{degen}
     \lim_{\la \rightarrow \pm 2 \i} \frac{R^{[1,1]} (\la)}{2 b(\la)}
        = P^\pm \epc \qd
     P^+ = \begin{pmatrix}
              1 &&& \\ & \tst{\2} & \tst{\2} & \\
	      & \tst{\2} & \tst{\2} & \\ &&& 1
           \end{pmatrix} \epc \qd
     P^- = \begin{pmatrix}
              0 &&& \\ & \mspace{14.mu} \tst{\2} & - \tst{\2} & \\
              & - \tst{\2} & \mspace{14.mu} \tst{\2} & \\ &&& 0
           \end{pmatrix} \epp
\end{equation}
The $P^\pm$ are the orthogonal projectors onto the singlet and triplet
subspaces $V^{(s)}, V^{(t)} \subset {\mathbb C}^2 \otimes {\mathbb C}^2$
with standard bases
\begin{align}
     B^{(s)} =
     & \Bigl\{ \tst{\frac{1}{\sqrt{2}}}
	     \Bigl(
             \bigl(\begin{smallmatrix} 1 \\ 0 \end{smallmatrix}\bigr)
	     \otimes
             \bigl(\begin{smallmatrix} 0 \\ 1 \end{smallmatrix}\bigr) -
             \bigl(\begin{smallmatrix} 0 \\ 1 \end{smallmatrix}\bigr)
	     \otimes
             \bigl(\begin{smallmatrix} 1 \\ 0 \end{smallmatrix}\bigr)
	     \Bigr) \Bigr\} \epc \qd \notag \\[1ex]
     B^{(t)} =
     & \Bigl\{
             \bigl(\begin{smallmatrix} 1 \\ 0 \end{smallmatrix}\bigr)
	     \otimes
             \bigl(\begin{smallmatrix} 1 \\ 0 \end{smallmatrix}\bigr),
             \tst{\frac{1}{\sqrt{2}}}
	     \Bigl(
             \bigl(\begin{smallmatrix} 1 \\ 0 \end{smallmatrix}\bigr)
	     \otimes
             \bigl(\begin{smallmatrix} 0 \\ 1 \end{smallmatrix}\bigr) +
             \bigl(\begin{smallmatrix} 0 \\ 1 \end{smallmatrix}\bigr)
	     \otimes
             \bigl(\begin{smallmatrix} 1 \\ 0 \end{smallmatrix}\bigr)
	     \Bigr),
             \bigl(\begin{smallmatrix} 0 \\ 1 \end{smallmatrix}\bigr)
	     \otimes
             \bigl(\begin{smallmatrix} 0 \\ 1 \end{smallmatrix}\bigr)
	     \Bigr\} \epp
\end{align}
Due to (\ref{ybe}) and (\ref{degen}) we have the important relation
\begin{equation} \label{tripinvar}
     P_{23}^- R^{[1,1]}_{13} (\la) R^{[1,1]}_{12} (\la + 2 \i) P_{23}^+ = 0
        \epc
\end{equation}
meaning that $R^{[1,1]}_{13} (\la) R^{[1,1]}_{12} (\la + 2 \i)$ leaves
${\mathbb C}^2 \otimes V^{(t)}$ invariant.

Let us introduce $U: {\mathbb C}^2 \otimes {\mathbb C}^2 \rightarrow
{\mathbb C}$ and $S: {\mathbb C}^2 \otimes {\mathbb C}^2 \rightarrow
{\mathbb C}^3$,
\begin{equation} \label{defus}
     U = \Bigl(0, \tst{\frac{1}{\sqrt{2}}}, - \tst{\frac{1}{\sqrt{2}}},
               0\Bigr) \epc \qd
     S = \begin{pmatrix}
            1 &&& \\ & \tst{\frac{1}{\sqrt{2}}} &
            \tst{\frac{1}{\sqrt{2}}} & \\ &&& 1
         \end{pmatrix}
\end{equation}
which map the singlet and triplet subspaces of the tensor product of
two spin-$\2$ representations onto ${\mathbb C}$ or ${\mathbb C}^3$,
respectively. These matrices satisfy
\begin{equation} \label{propus}
     S S^t = I_3 \epc \qd S^t S = P^+ \epc \qd
     U U^t = 1 \epc \qd U^t U = P^- \epc
\end{equation}
where the superscript $t$ indicates the transposition of matrices.

Using $S$ we can define the fused $R$-matrices
\begin{subequations}
\label{fusedr}
\begin{align}
     R^{[1,2]} (\la) & =
        S_{23}\, R_{13}^{[1,1]} (\la)
	         R_{12}^{[1,1]} (\la + 2\i)\, S_{23}^t \epc \\[1ex]
     R^{[2,1]} (\la) & =
        S_{12}\, R_{13}^{[1,1]} (\la - 2\i)
	         R_{23}^{[1,1]}\, (\la) S_{12}^t \epc \\[1ex]
     R^{[2,2]} (\la) & = S_{12} S_{34} R^{[1,1]}_{14} (\la - 2 \i)
                 R^{[1,1]}_{13} (\la) R^{[1,1]}_{24} (\la)
		 R^{[1,1]}_{23} (\la + 2 \i) S_{34}^t S_{12}^t
\end{align}
\end{subequations}
acting on ${\mathbb C}^2 \otimes {\mathbb C}^3$, ${\mathbb C}^3 \otimes
{\mathbb C}^2$, or ${\mathbb C}^3 \otimes {\mathbb C}^3$, respectively.
Combining the Yang-Baxter equation (\ref{ybe}) and equations
(\ref{tripinvar}), (\ref{propus}) it is easy to see that
\begin{equation} \label{ybes}
     R^{[2s_1,2s_2]}_{12} (\la - \m) R^{[2s_1,2s_3]}_{13} (\la)
        R^{[2s_2,2s_3]}_{23} (\m)
      = R^{[2s_2,2s_3]}_{23} (\m) R^{[2s_1,2s_3]}_{13} (\la)
        R^{[2s_1,2s_2]}_{12} (\la - \m) \epc
\end{equation}
where $s_j = \2, 1$ for $j = 1, 2, 3$.

In particular, $R^{[2,2]}$ is a solution of the Yang-Baxter equation.
With $P^{[2]}$ denoting the transposition on ${\mathbb C}^3 \otimes
{\mathbb C}^3$ and $\check R^{[2,2]} = P^{[2]} R^{[2,2]}$ it has the
further properties
\begin{subequations}
\label{regun}
\begin{align} \label{regular}
     & R^{[2,2]} (0) = P^{[2]} \epc \qd
      % P^{[2]} R^{[2,2]} (\la) P^{[2]} = R^{[2,2]} (\la)
       \\[1ex] \label{unitary}
     & \check R^{[2,2]} (\la)\, \check R^{[2,2]} (-\la) = I_9 \epc
\end{align}
\end{subequations}
i.e., $R^{[2,2]}$ is regular
%, symmetric with respect to the transposition of spaces
and unitary. It follows with (\ref{regular}) that $R^{[2,2]}$ generates
the Hamiltonian (\ref{ham}),
\begin{equation} \label{locham}
     H = \i J \sum_{n = - L + 1}^L h_{n-1, n} \epc \qd h_{n-1, n} =
         \6_\la \check R^{[2,2]}_{n-1, n} (\la) \bigr|_{\la = 0} \epp
\end{equation}
\subsection{Density matrix}
In \cite{GKS04a} we have set up a formalism which enables us to calculate
thermal correlation functions in integrable models with $R$-matrices
fulfilling (\ref{regular}). It is based on the so-called quantum transfer
matrix \cite{Suzuki85} and its associated monodromy matrix which are
directly related to the statistical operator.

The Hamiltonian (\ref{ham}) preserves the total spin
\begin{equation}
     S^\a = \sum_{j = - L + 1}^L S_j^\a \epp
\end{equation}
Thus, the magnetization in $z$-direction is a thermodynamic quantity,
and the statistical operator
\begin{equation}
     \r_L (T, h) = \re^{ - \frac{H - 2 h S^z}{T}}
\end{equation}
describes the spin chain (\ref{ham}) in thermal equilibrium at temperature
$T$ and magnetic field~$h$.

The statistical operator does not exist in the thermodynamic limit.
Quantities that are better defined for the infinite chain are the free
energy per lattice site and the density matrix of a finite chain segment.
The free energy per lattice site is
\begin{equation}
     f(T,h) = - T \lim_{L \rightarrow \infty}
              \frac{\ln \tr_{- L + 1, \dots, L}\, \r_L (T, h)}{2L} \epp
\end{equation}
It determines the thermodynamics of the model \cite{Suzuki99} which
will be briefly reviewed in section \ref{sec:therm}. The density matrix
of a finite chain segment $[1,m]$ is defined as
\begin{equation} \label{defdensmat}
     D_{[1,m]} (T, h) =
        \lim_{L \rightarrow \infty}
	\frac{\tr_{- L + 1, \dots, 0, m + 1, \dots, L}\, \r_L (T, h)}
	     {\tr_{- L + 1, \dots, L}\, \r_L (T, h)} \epp
\end{equation}
With $D_{[1,m]} (T, h)$ we can calculate the expectation value of any
local operator that acts trivially outside the finite segment $[1,m]$.
In particular, $D_{[1,m]} (T, h)$ allows us to calculate the static
correlation functions inside $[1,m]$.

For any integrable model, whose $R$-matrix does not only satisfy the
Yang-Baxter equation, but also the regularity and unitarity conditions
(\ref{regun}), we can approximate the statistical operator $\r_L (T, h)$
of the $2L$-site Hamiltonian using the monodromy matrix of an
appropriately defined vertex model with $2L$ vertical lines ($- L + 1,
\dots, L$) and $N$ alternating horizontal lines ($\overline 1, \dots,
\overline N$ with $N$ even). This fact was exploited many times in the
calculation of the bulk thermodynamic properties of integrable quantum
chains, in particular, in case of the higher-spin integrable Heisenberg
chains \cite{Suzuki99}. In \cite{GKS04a} it was noticed that the same
formalism is also useful for the calculation of thermal correlation
functions. Following the general prescription in \cite{GKS04a} we define
\begin{multline}
     T_j^{[2]} (\la) = \re^{2 h S_j^z/T}
        R_{j, \overline N}^{[2,2]} (\la - \be/N)
        R_{\overline{N-1}, j}^{[2,2] \: t_1} (-\be/N - \la) \dots \\ \dots
        R_{j, \overline 2}^{[2,2]} (\la - \be/N)
        R_{\overline 1, j}^{[2,2] \: t_1} (-\be/N - \la) \epc
\end{multline}
where $t_1$ indicates transposition with respect to the first space in a
tensor product. This monodromy matrix is constructed in such a way that
(see \cite{GKS04a})
\begin{equation} \label{piovertwo}
      \tr_{\bar 1 \dots \overline N} \Bigl\{
         T^{[2]}_{- L + 1} (0) \dots T^{[2]}_L (0) \Bigr\}
         = \biggl[1 - \frac{2}{NT} \sum_{n = - L + 1}^L
                      \bigr( \be T h_{n-1,n} - 2 h S_n^z \bigr)
		      + \CO \Bigl( \frac{1}{N^2} \Bigr)
                      \biggr]^\frac{N}{2} \mspace{-9.mu} .
\end{equation}
Hence, setting $\be = \i J/T$ and
\begin{equation}
     \r_{N, L} (T, h) =
      \tr_{\bar 1 \dots \overline N} \Bigl\{
         T^{[2]}_{- L + 1} (0) \dots T^{[2]}_L (0) \Bigr\} \epc
\end{equation}
we conclude, using (\ref{ham}), (\ref{locham}) and (\ref{piovertwo}),
that
\begin{equation}
     \lim_{N \rightarrow \infty} \r_{N, L} (T, h) = \r_{L} (T, h) \epp
\end{equation}
We shall call this limit the Trotter limit.

The transfer matrix
\begin{equation}
     t^{[2]} (\la ) = \tr_j T^{[2]}_j (\la )
\end{equation}
is commonly called the quantum transfer matrix. We shall recall below
how it can be diagonalized by means of the algebraic Bethe ansatz
\cite{Suzuki99}. Quite generally it has the remarkable property that
the eigenvalue $\La^{[2]} (0)$ of largest modulus of $t^{[2]} (0)$ (we
call it the dominant eigenvalue) is real and non-degenerate and is
separated by the rest of the spectrum by a gap \cite{Suzuki85,SuIn87}.
It can further be shown that
\begin{equation} \label{free}
     f (T,h) = - T \lim_{L \rightarrow \infty} \lim_{N \rightarrow \infty}
            \frac{\ln \tr_{- L + 1, \dots, L}\, \r_{N, L} (T, h)}{2 L}
       = - T \lim_{N \rightarrow \infty} \ln \La^{[2]} (0) \epp
\end{equation}
Thus, the dominant eigenvalue alone determines the bulk thermodynamic
properties of the spin chain.

Owing to the fact that $R^{[2,2]}$ satisfies the Yang-Baxter equation
the transfer matrices $t^{[2]} (\la)$ form a commutative family,
\begin{equation}
     [t^{[2]} (\la), t^{[2]} (\m)] = 0 \epp
\end{equation}
It follows that the eigenvectors of $t^{[2]} (\la)$ do not depend on
$\la$. Let $|\Ps_0\>$ denote an eigenvector belonging to the dominant
eigenvalue $\La^{[2]} (0)$. We shall call it the dominant eigenvector.
It is unique up to normalization and is an eigenvector of $t^{[2]} (\la)$
with eigenvalue $\La^{[2]} (\la) = \<\Ps_0| t^{[2]} (\la)|\Ps_0\>/
\<\Ps_0|\Ps_0\>$. In \cite{GKS04a} it was pointed out that such an
eigenvector determines all static correlation functions at temperature
$T$ and magnetic field $h$. In particular, it determines the density
matrix (\ref{defdensmat}) of any finite segment $[1,m]$,
\begin{equation} \label{d2hom}
     D_{[1,m]} (T, h) = \lim_{N \rightarrow \infty}
        \frac{\<\Ps_0| T^{[2]} (0) \otimes \dots \otimes
                       T^{[2]} (0) |\Ps_0\>}
             {\<\Ps_0|\Ps_0\> \bigl( \La^{[2]} (0)\bigr)^m} \epp
\end{equation}

For technical reasons it is better to consider a slightly more general
expression than the one under the limit, by allowing for mutually
distinct spectral parameters $\x_j$, $j = 1, \dots, m$, instead of zero.
Setting $\x = (\x_1, \dots, \x_m)$ we define
\begin{equation} \label{d2inhom}
     D^{[2]} (\x) = \frac{\<\Ps_0| T^{[2]} (\x_1) \otimes \dots
                          \otimes T^{[2]} (\x_m) |\Ps_0\>}
                         {\<\Ps_0|\Ps_0\> \La^{[2]} (\x_1) \dots
                          \La^{[2]} (\x_m)} \epc
\end{equation}
the inhomogeneous density matrix at finite Trotter number. Then
\begin{equation} \label{d2lim}
     D_{[1,m]} (T, h) = \lim_{N \rightarrow \infty} \:
                      \lim_{\x_1, \dots, \x_m \rightarrow 0}
                      D^{[2]} (\x) \epp
\end{equation}
The expression (\ref{d2inhom}) is our starting point for the derivation
of the multiple integral representation in appendix \ref{app:dermult}.
\subsection{Bethe Ansatz solution}
For the calculation of the free energy (\ref{free}) and the inhomogeneous
density matrix (\ref{d2inhom}) we need to know in first place the
dominant eigenvector $|\Ps_0\>$ and the corresponding transfer matrix
eigenvalue $\La^{[2]} (\la)$. They can be obtained by means of the
standard algebraic Bethe ansatz for the spin-$\2$ generalized model
(see e.g.\ chapter 12.1.6 of \cite{thebook}), since, by the general
reasoning of the fusion procedure \cite{KuSk82b}, the quantum transfer
matrix $t^{[2]} (\la)$ can be expressed in terms of a transfer matrix
with spin-$\2$ auxiliary space and its associated quantum determinant.

For the temperature case at hand we define the staggered monodromy
matrix with spin-$\2$ auxiliary space \cite{Suzuki99} by
\begin{multline} \label{defmono12}
     T_a^{[1]} (\la + \i) = \re^{h \s_a^z/T}
        R_{a, N}^{[1,2]} (\la - \be/N)
        R_{N-1, a}^{[2,1] \: t_1} (-\be/N - \la) \dots \\ \dots
        R_{a, 2}^{[1,2]} (\la - \be/N)
        R_{1, a}^{[2,1] \: t_1} (-\be/N - \la) \epp
\end{multline}
Then, interpreting this monodromy matrix as a $2 \times 2$ matrix in
the auxiliary space $a$, we define
\begin{equation} \label{deft1detq1}
     t^{[1]} (\la) = \tr T^{[1]} (\la) \epc \qd
     \detq T^{[1]} (\la) = U \bigl( T^{[1]} (\la - \i)
                               \otimes T^{[1]} (\la + \i) \bigr) U^t \epp
\end{equation}
It follows from (\ref{fusedr}) that
\begin{equation} \label{spin1mono}
     T^{[2]} (\la) = S \bigl( T^{[1]} (\la - \i)
                              \otimes T^{[1]} (\la + \i) \bigr) S^t \epp
\end{equation}
Taking the trace and using (\ref{deft1detq1}) we conclude that
\begin{equation} \label{t2t1qdet}
     t^{[2]} (\la) = t^{[1]} (\la - \i) t^{[1]} (\la + \i)
                     - \detq T^{[1]} (\la) \epc
\end{equation}
Hence, since $\detq T^{[1]} (\la)$ commutes with $T^{[1]} (\la)$ 
\cite{KuSk82b}, every eigenstate of $t^{[1]} (\la)$ is an eigenstate of
$t^{[2]} (\la)$ as well.

The algebraic Bethe ansatz is based on the Yang-Baxter algebra relations
\begin{equation} \label{yba}
     \check R^{[1,1]} (\la - \m)
        \bigl( T^{[1]} (\la) \otimes T^{[1]} (\m) \bigr) =
        \bigl( T^{[1]} (\m) \otimes T^{[1]} (\la) \bigr)
        \check R^{[1,1]} (\la - \m)
\end{equation}
which follow from (\ref{ybes}) and (\ref{defmono12}). Representing
$T^{[1]} (\la)$ by the $2 \times 2$ matrix
\begin{equation}
     T^{[1]} (\la) = \begin{pmatrix}
                        A(\la) & B(\la) \\ C(\la) & D(\la)
                     \end{pmatrix}
\end{equation}
and defining the pseudo vacuum
\begin{equation}
     |0\> =
        \Bigl[
        \Bigl( \begin{smallmatrix} 0 \\ 0 \\ 1 \end{smallmatrix} \Bigr)
        \otimes
        \Bigl( \begin{smallmatrix} 1 \\ 0 \\ 0 \end{smallmatrix} \Bigr)
        \Bigr]^{\otimes \frac{N}{2}}
\end{equation}
we deduce from (\ref{defmono12}) that
\begin{equation}
     C(\la) |0\> = 0 \epc \qd
     A(\la) |0\> = a(\la) |0\> \epc \qd D(\la) |0\> = d(\la) |0\> \epc
\end{equation}
where the pseudo vacuum eigenvalues $a(\la)$ and $d(\la)$ are explicit
complex valued functions. Using the notation
\begin{equation} \label{defphi}
     \phi_\pm (\la) = (\la \pm \i u)^{N/2} \epc \qd u = - \frac{J}{NT}
\end{equation}
which proved to be useful in \cite{Suzuki99}, we can express them as
\begin{equation} \label{vacexp}
     a(\la) = \frac{\re^{h/T} \phi_- (\la + \i)}{\phi_- (\la - 3 \i)} \epc
        \qd
     d(\la) = \frac{\re^{-h/T} \phi_+ (\la - \i)}{\phi_+ (\la + 3 \i)} \epp
\end{equation}

Given the Yang-Baxter algebra (\ref{yba}) and the pseudo vacuum
eigenvalues (\ref{vacexp}) the eigenvectors and eigenvalues of
$t^{[1]} (\la)$ can be obtained from general considerations (see e.g.\
chapter 12.1.6 of \cite{thebook}). The dominant eigenstate $|\Ps_0\>$
of $t^{[2]} (\la)$, in particular, can be represented as
\begin{equation} \label{domi}
     |\Ps_0\> = B(\la_1) \dots B(\la_N) |0\> \epc
\end{equation}
where the set of so-called Bethe roots $\{\la_j\}_{j=1}^N$ is a specific
solution of the Bethe ansatz equations
\begin{equation} \label{bae}
      \frac{a(\la_j)}{d(\la_j)} = \prod_{\substack{k = 1 \\ k \ne j}}^N
         \frac{\la_j - \la_k + 2\i}{\la_j - \la_k - 2\i} \epc \qd
         j = 1, \dots, N \epp
\end{equation}
For the given set of Bethe roots $\{\la_j\}_{j=1}^N$ we define the
$Q$-function
\begin{equation} \label{defq}
     q(\la) = \prod_{j=1}^N (\la - \la_j) \epp
\end{equation}
Then the eigenvalue of $t^{[1]} (\la)$ corresponding to $|\Ps_0\>$ is
\begin{equation}
     \La^{[1]} (\la) = a(\la) \frac{q(\la - 2 \i)}{q(\la)}
          + d(\la) \frac{q(\la + 2 \i)}{q(\la)} \epp
\end{equation}
As for the eigenvalue of $t^{[2]} (\la)$ we conclude with (\ref{t2t1qdet})
and equation (\ref{simpleqdet}) below that
\begin{equation} \label{eva}
     \La^{[2]} (\la) = \La^{[1]} (\la - \i) \La^{[1]} (\la + \i)
                         - a(\la + \i) d(\la - \i)
\end{equation}
This eigenvalue and the Bethe ansatz equations (\ref{bae}) are the main
input for the calculation of the thermodynamics of the spin-1
chain. In order to perform the Trotter limit the eigenvalue must
be represented by means of auxiliary functions satisfying a finite
set of nonlinear integral equations. This was achieved in
\cite{Suzuki99}. To the extend we need the results also for the
calculation of the density matrix, they are reviewed in the following
section.
\subsection{Simplified form of fused monodromy matrix}
Slight simplifications are possible for the form (\ref{spin1mono}) of the
fused monodromy matrix $T^{[2]}$ and for the form (\ref{deft1detq1}) of
the quantum determinant of $T^{[1]}$. We include them here for later
convenience. Setting $T^\pm = T^{[1]} (\x \pm \i)$,
\begin{equation}
     T^\pm = \begin{pmatrix} A^\pm & B^\pm \\ C^\pm & D^\pm
             \end{pmatrix}
\end{equation}
and using the Yang-Baxter equation and (\ref{propus}), we conclude that
\begin{equation} \label{monolr}
      T^{[2]} (\x) = S(T^+ \otimes T^-)S^t =
                     S(T^- \otimes T^+)S^t \epp
\end{equation}
Similarly, it follows that
\begin{equation}
      S(T^+ \otimes T^-)U^t = U(T^- \otimes T^+)S^t = 0
\end{equation}
with the help of which we can represent $T^{[2]}$ e.g.\ as
\begin{equation} \label{t1form}
     T^{[2]} (\x) =
        \begin{pmatrix}
	   A^- A^+ & \sqrt{2} A^+ B^- & B^+ B^- \\
	   \sqrt{2} C^- A^+ & C^- B^+ + D^- A^+ & \sqrt{2} D^- B^+ \\
	   C^- C^+ & \sqrt{2} C^+ D^- & D^- D^+
        \end{pmatrix}
\end{equation}
and $\detq T^{[1]} (\la)$ as
\begin{equation} \label{simpleqdet}
     \detq T^{[1]} (\x) = D^- A^+ - B^- C^+ \epp
\end{equation}
%=========================================================================
\section{Thermodynamics}
\label{sec:therm}
In this section we consider the evaluation of the free energy per lattice
site $f(T,h)$ by means of nonlinear integral equations (NLIE). This gives
us the opportunity to introduce certain auxiliary functions and
integration contours that are also relevant for the multiple integral
representation of the density matrix elements in the next section. Our
starting point is the expression (\ref{free}) for $f(T,h)$ in terms of the
dominant eigenvalue $\La^{[2]} (\la)$ of the quantum transfer matrix
together with the Bethe ansatz solution (\ref{bae})-(\ref{eva}). In
\cite{Suzuki99} the problem was solved within the more general context of
the fusion hierarchy, and NLIE for the integrable isotropic spin chains of
arbitrary spin were obtained. We believe that those NLIE are optimal in
several respects for the calculation of the free energy. They are integral
equations of convolution type formulated for a minimal number of functions
on straight lines, and, for this reason, can be accurately solved
numerically. Moreover, the low temperature asymptotics of the free energy
can be extracted from these equations \cite{Suzuki99}.

For the calculation of the free energy for spin 1 we will be dealing with
three coupled NLIE for three functions $\fb$, $\fbq$ and $y$. We show the
equations below in (\ref{nlie}) and present an alternative derivation in
appendix~\ref{app:contnlietonlie}. For finite Trotter number $N$ the
functions $\fb$, $\fbq$ and $y$ can be expressed in terms of the
$Q$-functions (\ref{defq}) and the functions $\Ph_\pm$ introduced in
(\ref{defphi}) (see appendix~\ref{app:auxfun}). This defines them as
meromorphic functions in the entire complex plane, but is inappropriate
for performing the Trotter limit. In the NLIE, on the other hand, the
Trotter number appears only in the driving term and the Trotter limit
is easily obtained. For a discussion of some of the subtleties related
to the Trotter limit and the definition of useful auxiliary functions
see \cite{GoSu10}.

If one is only interested in the free energy, it is sufficient to know
the functions $\fb$, $\fbq$ and $y$ close to the real axis (see
(\ref{nlie}), (\ref{evaint}) below). For the calculation of more general
physical quantities, however, as, for instance, the density matrix
elements we are going to consider in the next section, we need to know
$\fb$, $\fbq$ also close to straight lines parallel to the real axis,
passing through $\pm 2 \i$. This is the reason why we reconsider and
slightly extend the approach of \cite{Suzuki99}.

\begin{figure}
\begin{center}
\includegraphics[height=17em]{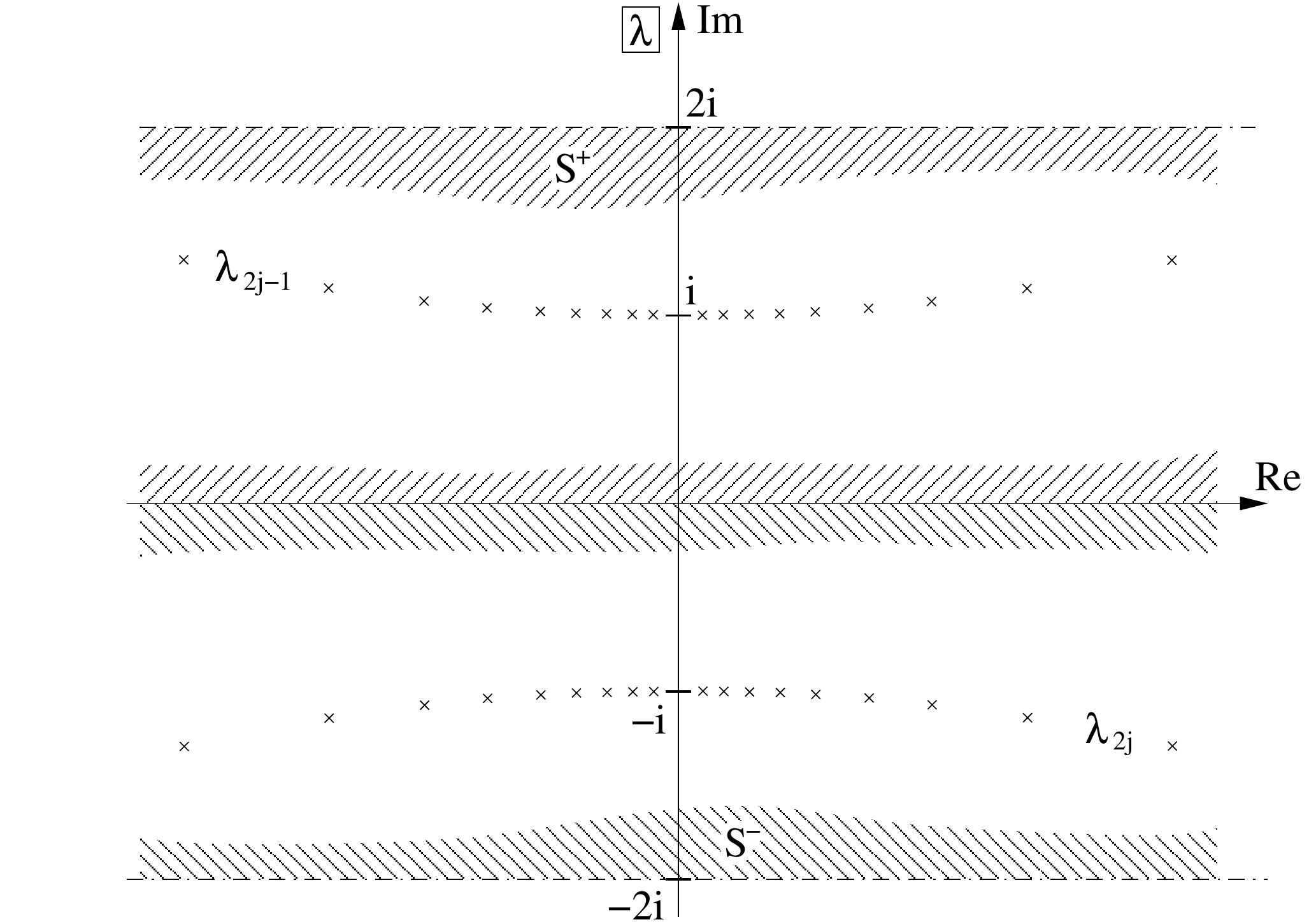} 
\caption{\label{fig:stripsandroots} Schematic distribution of the upper
and lower Bethe roots $\la_{2j-1}$ and $\la_{2j}$, respectively, in the
strips $\CS^\pm$.}
\end{center}
\end{figure}

The necessity of considering auxiliary functions in an extended strip
around the real axis originates from the particular distribution of
the Bethe roots that parameterize the dominant state. Define the
strips
\begin{equation}
     \CS^\pm = \bigl\{\la \in {\mathbb C} \big|
                      0 < \pm \Im \la < 2 \bigr\} \epp
\end{equation}
Then the Bethe roots of the dominant state come in $N/2$ pairs (so-called
two-strings) with one root in $\CS^+$ and the other one in $\CS^-$.
For large Trotter number they accumulate in the vicinity of $\pm \i$.
We shall call the Bethe roots in $\CS^+$ the upper Bethe roots and the
Bethe roots in $\CS^-$ the lower Bethe roots. By convention the upper
Bethe roots will be denoted $\la_{2j - 1}$ and the lower Bethe roots
$\la_{2j}$, where $j = 1, \dots, N/2$ (see figure~\ref{fig:stripsandroots}).

%=======================================================================%
%=======================================================================%

Typical physical quantities at finite temperature can be written as
sums over the Bethe roots of the dominant state. Such sums can be converted
into contour integrals by means of appropriate auxiliary functions
having their zeros at the Bethe roots. As compared to the spin-$\2$ case the
choice of the contours and auxiliary functions is more delicate for spin 1.
In particular, it seems that the auxiliary functions and integration
contours have to be chosen separately in $\CS^+$ and $\CS^-$. We shall
consider the auxiliary functions
\begin{equation} \label{deff}
     \ff (\la) = \frac{1}{\fb (\la - 2 \i)} \epc \qd
     \ffq (\la) = \frac{1}{\fbq (\la + 2 \i)}
\end{equation}
(see appendix~\ref{app:auxfun} for the definitions of $\fb$ and $\fbq$
in terms of $Q$-functions). As usually we also introduce the corresponding
`capital functions'
\begin{equation}
     \Ff (\la) = 1 + \ff (\la) \epc \qd \Ffq (\la) = 1 + \ffq (\la) \epp
\end{equation}
They are meromorphic for finite Trotter number, and $\Ff$ has in
$\CS^+$ exactly $N/2$ zeros located at the upper Bethe roots and
only a single $N/2$-fold pole at $\i - \i u$. Similarly, $\Ffq$
has in $\CS^-$ exactly $N/2$ zeros located at the lower Bethe roots and
only a single  $N/2$-fold pole at $- \i + \i u$.

%The machineries of the integrable system show that they are parameterized
%by the Bethe  ansatz roots $\{\la_j\}$. The formal expressions with
%$\{\la_j\}$, however meet serious problems in the Trotter limit, as
%infinitely many of $\{\la_j\}$ accumulate near the origin. See
%\cite{GoSu10} for a discussion related to this.
% 
%In such circumstances, the lesson from the spin-$\frac{1}{2}$ case suggests
%that it is better to rewrite the summation over $\{\la_j\}$ by contour
%integrals using appropriate auxiliary function(s). The difference from
%the spin $\frac{1}{2}$ is that $\{\la_j\}$ forms the two-string
%configuration for the spin-1 case in the ground state. The roots are thus
%widely separated in the vertical direction. A contour encircling
%$\{\la_j\}$ can contain the irrelevant contributions (zeros or poles of
%auxiliary functions) which will ruin the equivalence of the summation
%expression and the integral expression. We therefore have to carefully
%choose auxiliary functions and integration contours, in order to establish
%the  desired expression. 

Using this information and the definitions of some additional useful
auxiliary functions in terms of $Q$-functions (see
appendix~\ref{app:auxfun}) we obtain the following NLIE,
\begin{subequations}
\label{LCN}
\begin{align}
     \ln \frac{\ff(\la)}{\afIIq (\la)} & =
        d_f(\la) +\ln \frac{\Ff(\la)}{\Bfq(\la) } \notag \\
     & + \int_{\CC^+} \frac{\rd \m}{2 \p \i} \jg2(\la-\mu) \ln \Ff (\mu)
       + \int_{\CCq^-} \frac{\rd \m}{2 \p \i} \jg2(\la-\mu) \ln \Ffq (\mu)
         \epc \quad \la \in \CC^+ \epc \label{LCN1} \\
     \ln \frac{\ffq (\la)}{\afII(\la)}
     & = - d_f(\la) + \ln \frac{\Ffq(\la)}{\Bf(\la)} \notag \\
     & - \int_{\CCq^+} \frac{\rd \m}{2 \p \i} \jg2(\la-\mu) \ln \Ff (\mu)
       - \int_{\CC^-} \frac{\rd \m}{2 \p \i} \jg2(\la-\mu) \ln \Ffq (\mu)
         \epc \quad \la \in \CC^- \epp \label{LCN2}
\end{align}
\end{subequations}
Here we have introduced the kernel
\begin{equation}
       \jg2(\la) = \frac{1}{\la-2 \i} - \frac{1}{\la+2 \i}
\end{equation}
and the driving term
\begin{multline}
     d_f(\la) = \frac{2 h}{T}
                + \ln \frac{\phi_+(\la-3\i) \phi_-(\la-\i)
		            \phi_-(\la+\i) \phi_+(\la+3\i)}
			   {\phi_-(\la-3\i) \phi_+(\la-\i)
			    \phi_+(\la+\i) \phi_-(\la+3\i)} \\[1ex]
       \overset{N \rightarrow \infty}{\longrightarrow}\ \frac{2 h}{T}
       + \frac{\i J}{T} \bigl( \jg2(\la+\i)-\jg2(\la-\i) \bigr) \epp
\end{multline}
Note that the only explicit $N$-dependence is in the driving term
$d_f$. And since this driving term has a simple Trotter limit, we conclude
that the functions $\ff$ and $\ffq$ have a Trotter limit as well.

%=======================================================================%
\begin{figure}
\begin{center}
\includegraphics[height=17em]{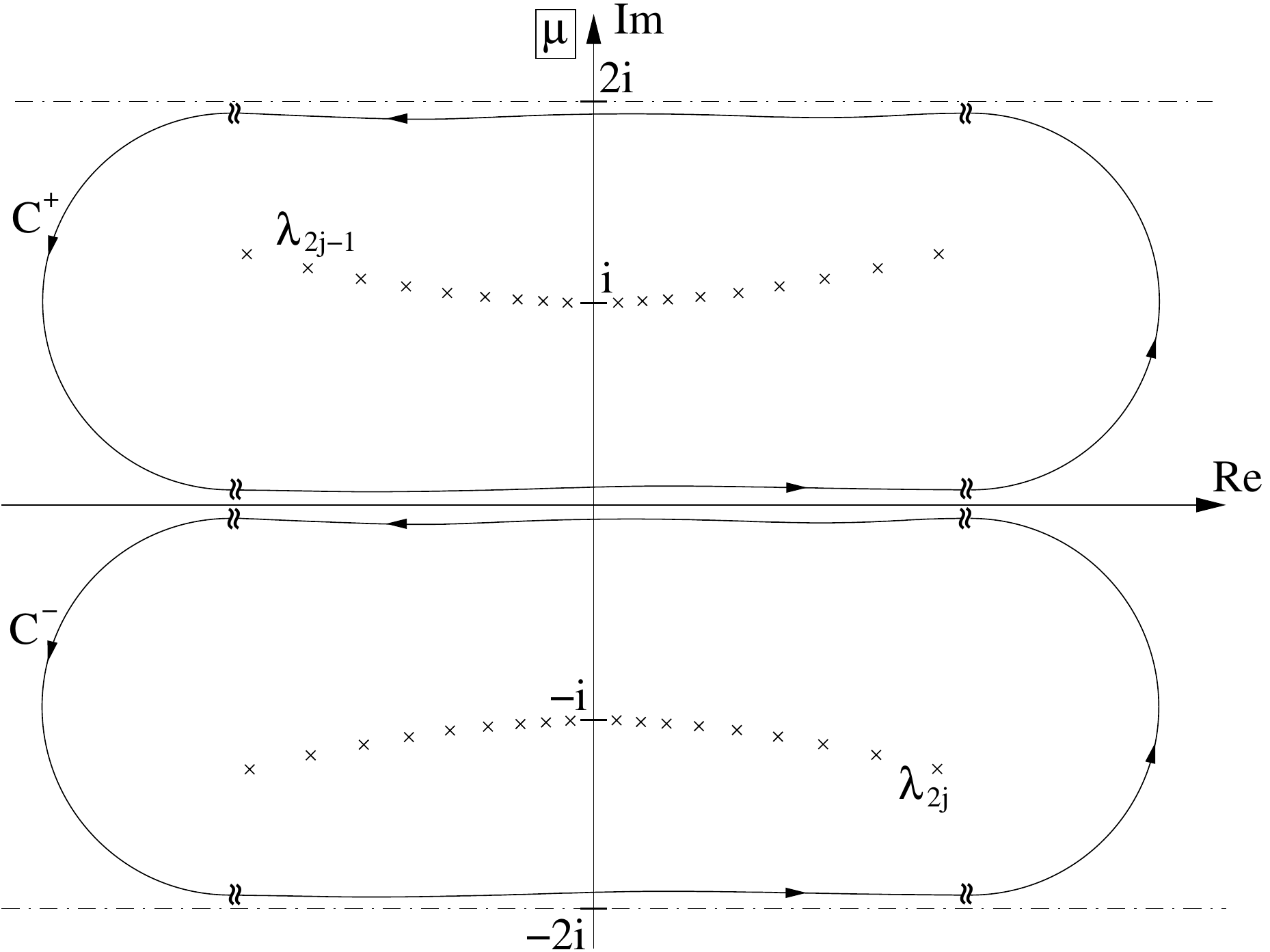} 
\caption{\label{fig:spin1contours} The contours $\CC^\pm$ in the strips
$\CS^\pm$ encircle the upper and lower Bethe roots respectively and close
at infinity.}
\end{center}
\end{figure}
%=======================================================================%

\begin{figure}
\begin{center}
\includegraphics[height=12em]{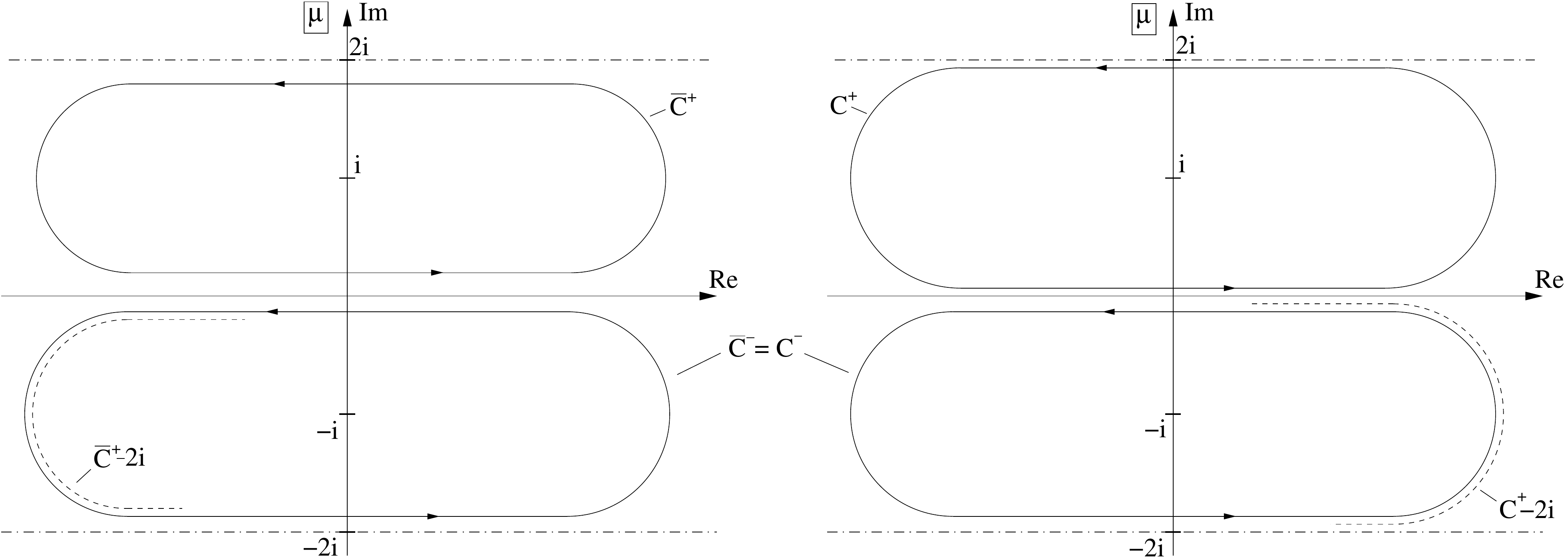} 
\caption{\label{fig:spin1narrower} For the regularization in the multiple
integral representation the dashed lines show the relative positions of
the contours $\CC^\pm$, $\CCq^\pm$.}
\end{center}
\end{figure}

The precise definition of the integration contours is slightly subtle.
We illustrate it in figure~\ref{fig:spin1contours}. $\CC^+$ is a simple
closed contour inside $\CS^+$ that encircles the upper Bethe roots.
We may realize it as a large rectangle with upper edge slightly below
$2 \i$ and lower edge slightly above the real axis. Similarly $\CC^-$
must enclose the lower Bethe roots inside $\CS^-$ and may also be
taken as a large rectangle, now with lower edge slightly above $-2 \i$
and with upper edge slightly below the real axis. The bar in  $\CCq^{\pm}$
means that the contours do not encircle the singularities originating
from the kernel $\jg2(\la)$. This prescription may be seen as an
`$\i \e$-regularization' of the kernel after the contour integral is
decomposed into an integral over straight lines. Such type of
regularization is needed because the kernel has poles at $\m =
\la \pm 2 \i$ which must not lie on the contours. Having in mind the
multiple integral representation in the next section we prefer to 
realize it in the way sketched in figure~\ref{fig:spin1narrower}, where
$\CCq^+ - 2\i$ inside $\CCq^- = \CC^-$ inside $\CC^+ - 2\i$, and `inside'
means `infinitesimally narrower'.

At first sight, (\ref{LCN1}) and (\ref{LCN2}) do not seem to be
enough to fix the unknown functions, as the number of equations is 
smaller than that of the functions. In order to understand that they
actually fix the functions $\ff$ and $\ffq$, let us simulate one step
in the iterative scheme. Assume that an approximate estimation of
$\ff, \ffq$ is already known. Then  $\afII$, $\afIIq$ are determined
from $\ff, \ffq$ by
\begin{equation}
     \afII (\la) = \frac{\frac{1}{\ff(\la+2\i)} -\ffq(\la)}{\Ffq(\la)}
                     \epc\ \text{for $\la \in \CC^-$,} \qd
     \afIIq(\la) = \frac{\frac{1}{\ffq(\la-2\i)} -\ff(\la)}{\Ff(\la)}
                     \epc\ \text{for $\la \in \CC^+$.}
\end{equation}
%\begin{subequations}
%\begin{align}
     %\afII (\la) & = \frac{\frac{1}{\ff(\la+2i)} -\ffq(\la)}{\Ffq(\la)}
                     %\qquad \la \in \CC^- \epc \\
     %\afIIq(\la) & = \frac{\frac{1}{\ffq(\la-2i)} -\ff(\la)}{\Ff(\la)}
                     %\qquad \la \in \CC^+ \epp
%\end{align}
%\end{subequations}
%=======================================================================%
%=======================================================================%

Note that  ${\Bfq(\la) }/{\Ff(\la)}$ and  ${\Bf(\la) }/{\Ffq(\la)}$ are
equal to $1+\afIIq(\la)$ and  $1+\afII(\la)$, respectively. They are thus
determined by given $\ff, \ffq$. Substituting them into the rhs of
(\ref{LCN1}) and (\ref{LCN2}) (and $\afII,\afIIq$ into the lhs), we obtain
the next-step approximation to $\ff, \ffq$. Therefore equations
(\ref{LCN}) consistently fix $\ff$ and $\ffq$. The other functions are
then determined from them.

Suppose that we have evaluated the auxiliary functions through
(\ref{LCN}). Then, for $|\Im \la| < 1$, the largest eigenvalue
$ \La^{[2]} (\la)$ is obtained as
\begin{align}
     \ln\La^{[2]}(\la) & = \ln \La^{[2]}_0(\la) +
        \int_{\CC^-} \frac{\rd \m}{2 \p \i}
        \jg2 (\la-\mu-3\i) \ln \Bf(\mu) +
	\int_{\CC^+} \frac{\rd \m}{2 \p \i}
        \jg2 (\la-\mu+\i) \ln \Bfq(\mu) \epc \label{La21} \notag \\[1ex]
     \ln \La^{[2]}_0 (\la) & = \frac{2 h}{T}
        + \ln \frac{\phi_+(\la-2\i) \phi_-(\la+2\i)}
	           {\phi_-(\la-2\i) \phi_+(\la+2\i)}\
          \overset{N \rightarrow \infty}{\longrightarrow}\
          \frac{2 h}{T} + \frac{4J}{T} \frac{1}{\la^2+4} \epp
\end{align}

The NLIE (\ref{LCN}) are actually only one of many possible choices. We
choose this one as we think that it has an advantage compared
to others in the following sense. Although the equations themselves
are literally correct, the integrations over contours suffer from poor
numerical accuracy, especially in the low temperature regime.
Therefore it is better to rewrite them in the form obtained in
\cite{Suzuki99}, where the integrations are defined on the straight lines.
We will show in appendix \ref{app:contnlietonlie} that (\ref{LCN}) can
be transformed into (\ref{nlie}) below with the help of additional
algebraic relations among the auxiliary functions. In the same appendix
\ref{app:contnlietonlie} we also provide subsidiary equations that
determine the functions $\ff$, $\ffq$ on straight lines close to the real
axis, which amounts to knowing $\fb$ and $\fbq$ on straight lines close
to $\pm 2\i$ (see (\ref{deff})).

Unlike in (\ref{LCN}) we need to deal with $\fb(\la), \fbq(\la)$
and $y(\la)$, if we choose straight lines as integration contours.
For convenience we introduce the shifted functions
\begin{equation}
    \fb_{\epsilon}(\la)=\fb(\la-\i\epsilon) \epc \quad
    \fbq_{\epsilon}(\la)=\fbq(\la+\i\epsilon) \epc
\end{equation}
and similar capital functions. Then the desired NLIE read
\begin{equation} \label{nlie}
     \begin{pmatrix}
        \log y(\la) \\
	\log \fb_{\epsilon} (\la) \\
	\log \fbq_{\epsilon}(\la)
     \end{pmatrix} = 
     \begin{pmatrix}
        0 \\
	\Delta_b(\la) \\
	\Delta_{\overline{b}}(\la)
     \end{pmatrix} +
     \widehat{\cal K} * 
     \begin{pmatrix}
        \log Y(\la) \\
	\log \Bf_{\epsilon} (\la) \\
	\log \Bfq_{\epsilon} (\la)
     \end{pmatrix} \epc
\end{equation}
where $(\widehat{\cal K}*g)_i$ denotes the matrix convolution
$\sum_j \int_{-\infty}^{\infty} \rd \m\, \widehat{\cal K}_{i,j}(\la-\mu)
g_j (\m)$, and
\begin{subequations}
\begin{align}
     \Delta_b(\la)
        & = - \frac{h}{T} + d(u,\la - \i \eps) \epc \qd
	  \Delta_{\overline{b}}(\la) = \frac{h}{T} + d(u,\la + \i \eps)
	  \epc \\[1ex]
     d(u,\la) & = \frac{N}{2}
        \int_{-\infty}^{\infty} \rd k\, \re^{- \i k \la}
	\frac{\sinh uk}{k\cosh k}\
	\overset{N \rightarrow \infty}{\longrightarrow}\
	- \frac{J}{T} \frac{\p}{2 \cosh \p \la /2} \epp
\end{align}
\end{subequations}
The integration constants ($\pm h/T$) are fixed by comparing the
asymptotic values of both sides of (\ref{nlie}) for $|\la| \rightarrow
\infty$. The kernel matrix is given by
\begin{equation} \label{ksym}
     \widehat{\cal {\cal K}}(\la) =
        \begin{pmatrix}
	   0 & {\cal K}(\la+\i\epsilon) & {\cal K}(\la-\i\epsilon) \\
	   {\cal K}(\la-\i\epsilon) & {\cal F}(\la) &
           -{\cal F}(\la+2\i(1-\epsilon)) \\
           {\cal K}(\la+\i\epsilon) & -{\cal F}(\la-2\i(1-\epsilon)) &
           {\cal F}(\la)
        \end{pmatrix} \epc
\end{equation}
where  
\begin{equation}
     {\cal K}(\la) = \frac{1}{4 \cosh\pi \la/2} \epc \qd
     {\cal F}(\la) = \int_{-\infty}^{\infty} \frac{\rd k}{2 \p}\,
                 \frac{e^{-|k|- \i k\la}}{2 \cosh k} \epp 
\end{equation}
The free energy then follows from (\ref{free}) noticing that the
dominant eigenvalue can be represented by integration over straight lines
as
\begin{multline} \label{evaint}
    \ln \Lambda^{[2]}(\la) = \ln \Lambda^{[2]}_0(\la) - \frac{2h}{T}\\
       + \int_{-\infty}^{\infty} \rd \m\; {\cal K}(\la - \m + \i \e)
         \ln \Bf_{\epsilon}(\mu)
       + \int_{-\infty}^{\infty} \rd \m\; {\cal K}(\la - \m - \i \e)
         \ln \Bfq_{\epsilon} (\mu) \epp
\end{multline}
As the actual transformation from (\ref{LCN}) to (\ref{nlie}) is
involved, we defer the details to appendix~\ref{app:contnlietonlie}.
%=========================================================================
\section{The multiple integral representation}
\label{sec:mult}
In this section we present the main result of this work, which is a
multiple integral formula for the matrix elements
${D^{[2]}}^{\a_1, \dots, \a_m}_{\be_1, \dots, \be_m} (\x)$, $\a_j, \be_k
= -, 0, +$, of the inhomogeneous density matrix (\ref{d2inhom}). Our
formula generalizes the result of \cite{Kitanine01} to finite temperature
and magnetic field and the result of \cite{GKS05} to spin 1. The details
of the derivation can be found in appendix~\ref{app:dermult}.

For any two sequences $(\a) = (\a_n)_{n=1}^m$ and $(\be) =
(\be_n)_{n=1}^m$ of upper and lower matrix indices we shall obtain a
different multiple integral. Let us introduce the notation $n_\s (x)$,
$\s = -, 0, +$, $(x) = (\a), (\be)$, for the number of $\s$s in the
sequence $(x)$, e.g. $n_0 (\be)$ is the number of zeros in $(\be)$. Then
\begin{subequations}
\begin{align}
     & n_+ (\a) + n_0 (\a) + n_- (\a) = m \epc \\
     & n_+ (\be) + n_0 (\be) + n_- (\be) = m \epc \\
     & n_+ (\be) - n_- (\be) - n_+ (\a) + n_- (\a) = 0 \epp
\end{align}
\end{subequations}
Here the last equation is equivalent to $2 n_+ (\a) + n_0 (\a) =
2 n_+ (\be) + n_0 (\be)$\footnote{Using (\ref{t1form}) this translates
into the fact that number of plus signs in the sequences of upper and
lower indices of the matrices $T^{[1]}$, the density matrix element
(\ref{d2symb}) is composed of, must be the same.}.

The dependence of the multiple integral on the indices $\a_j$, $\be_k$
enters through a sequence $(z) = (z_n)_{n=1}^{2m}$ encoding the positions
of $-, 0, +$ in $(\a)$ and $(\be)$. For the construction of $(z)$ we
order the density matrix indices as $\a_m, \dots, \a_1, \be_1, \dots,
\be_m$ and inspect them starting from the left. If $\a_m = -$ we do
nothing, if $\a_m = 0$ we define $z_1 = m$, and if $\a_m = +$ we
define $z_1 = z_2 = m$. We continue this procedure with $\a_{m-1}$
and so on. When we have reached $\a_1$ we have defined
\begin{equation}
     p = 2 n_+ (\a) + n_0 (\a)
\end{equation}
elements of the sequence $(z)$ in this way. If $\be_1 = -$ we define
$z_{p+1} = z_{p+2} = 1$, if $\be_1 = 0$ we define $z_{p+1} = 1$, and
if $\be_1 = +$ we do nothing. We continue the same way with $\be_2$,
$\be_3$ etc.\ until we end at $\be_m$. The sequence $(z)$ thus constructed
has $2 n_+ (\a) + n_0 (\a) + n_0 (\be) + 2 n_- (\be) = 2m$ elements,
and the pair $(z)$, $p$ is in one-to-one correspondence with the
sequences $(\a)$ and $(\be)$. As an example let us consider $(\a) =
(+, -, 0), \be = (0, 0, 0)$. Then $z_1 = 3$, $z_2 = z_3 = z_4 = 1$,
$z_5 = 2$, $z_6 = 3$, $p = 3$.

Two types of functions occur under the multiple integral. One type is
explicit and has its origin in the Yang-Baxter algebra.
The functions
\begin{subequations}
\label{d2wave}
\begin{align}
     F_\ell (\la) & = \prod_{k=1}^m (\la - \x_k - \i) \:
                      \prod_{k=1}^{\ell - 1} (\la - \x_k - 3 \i)
		      \prod_{k = \ell + 1}^m (\la - \x_k + \i) \epc \\
     \overline{F}_\ell (\la) & = \prod_{k=1}^m (\la - \x_k + \i) \:
                      \prod_{k=1}^{\ell - 1} (\la - \x_k + 3 \i)
		      \prod_{k = \ell + 1}^m (\la - \x_k - \i)
\end{align}
\end{subequations}
belong to this type. We think of them as `fused wave functions'.

The other type is related to the task of rewriting sums over Bethe roots
as integrals over closed contours (see appendix~\ref{app:dermult}). These
functions may be defined as solutions of linear integral equations over
closed contours. We have two pairs of such functions. The first one is
defined by
\begin{subequations}
\label{gpm}
\begin{align}
     & G^+ (\la, \x) =
        K(\la - \x - 3 \i) - K(\la - \x - \i)
        \notag \\ & \mspace{18.mu}
        - \int_{\CC^+} \frac{\rd \m}{2 \p \i}
             \frac{\Ff (\m)}{\Bfq (\m)} G^+ (\m, \x) K(\la - \m)
        + \int_{\CCq^-} \frac{\rd \m}{2 \p \i}
             \frac{\Ffq (\m)}{\Bf (\m)} G^- (\m, \x) K(\la - \m - 4\i)
             \epc \\[2ex]
     & G^- (\la, \x) =
        K(\la - \x + \i) - K(\la - \x + 3\i)
        \notag \\ & \mspace{18.mu}
        - \int_{\CCq^+} \frac{\rd \m}{2 \p \i}
             \frac{\Ff (\m)}{\Bfq (\m)} G^+ (\m, \x) K(\la - \m + 4\i)
        + \int_{\CC^-} \frac{\rd \m}{2 \p \i}
             \frac{\Ffq (\m)}{\Bf (\m)} G^- (\m, \x) K(\la - \m) \epc 
\end{align}
\end{subequations}
where $\la \in \CC^+$ for $G^+$ and $\la \in \CC^-$ for $G^-$.
The second pair of auxiliary functions needed in the definition of the
multiple integral is
\begin{subequations}
\label{spm}
\begin{align}
     & S^+ (\la, \x) =
        - \re(\la - \x - 5 \i) - \re(\la - \x - \i) - \frac{1}{Y (\x)}
        \bigl( K(\la - \x - 3 \i) + K (\la - \x - \i) \bigr)
        \notag \\ & \mspace{18.mu}
        - \int_{\CC^+} \frac{\rd \m}{2 \p \i}
             \frac{\Ff (\m)}{\Bfq (\m)} S^+ (\m, \x) K(\la - \m)
        + \int_{\CCq^-} \frac{\rd \m}{2 \p \i}
             \frac{\Ffq (\m)}{\Bf (\m)} S^- (\m, \x) K(\la - \m - 4\i)
             \epc \displaybreak[0] \\[2ex]
     & S^- (\la, \x) =
        - \re(\la - \x - \i) - \re(\la - \x + 3\i) - \frac{1}{Y (\x)}
        \bigl( K(\la - \x + \i) + K (\la - \x + 3\i) \bigr)
        \notag \\ & \mspace{18.mu}
        - \int_{\CCq^+} \frac{\rd \m}{2 \p \i}
             \frac{\Ff (\m)}{\Bfq (\m)} S^+ (\m, \x) K(\la - \m + 4\i)
        + \int_{\CC^-} \frac{\rd \m}{2 \p \i}
             \frac{\Ffq (\m)}{\Bf (\m)} S^- (\m, \x) K(\la - \m) \epc
\end{align}
\end{subequations}
where, similar to the above case, $\la \in \CC^+$ for $S^+$ and
$\la \in \CC^-$ for $S^-$ and where we have introduced the `bare energy
function'
\begin{equation}
     \re (\la) = \frac{1}{\la} - \frac{1}{\la + 2\i} \epp
\end{equation}

The functions $G^\pm$ and $S^\pm$ enter the multiple integral through
the determinant of a matrix with elements $\Th_{j, k}^{(p)}$ defined by
\begin{subequations}
\label{matrixtheta}
\begin{align}
     \Th_{j, 2k - 1}^{(p)} & =
          \begin{cases}
             G^+ (\om_j, \x_k) & j = 1, \dots, p \\
             G^- (\om_j, \x_k) & j = p + 1, \dots, 2m \epc
          \end{cases} \displaybreak[0] \\[2ex]
     \Th_{j, 2k}^{(p)} & =
          \begin{cases}
             \i S^+ (\om_j, \x_k) & j = 1, \dots, p \\
             \i S^- (\om_j, \x_k) & j = p + 1, \dots, 2m \epp
          \end{cases}
\end{align}
\end{subequations}

Using all of the above defined notation we can write the non-vanishing
matrix elements of the inhomogeneous spin-1 density matrix as
\begin{multline} \label{d2}
     {D^{[2]}}^{\a_1, \dots, \a_m}_{\be_1, \dots, \be_m} (\x) =
           \frac{2^{- m - n_+ (\a) - n_- (\be)}}
                {\prod_{1 \le j < k \le m} (\x_k - \x_j)^2
                 [(\x_k - \x_j)^2 + 4]} \\[1ex]
     \biggl[ \prod_{j=1}^p
             \int_{\cal C} \frac{\rd \om_j}{2 \p \i}
             F_{z_j} (\om_j) \biggr]
     \biggl[ \prod_{j = p + 1}^{2m} 
             \int_{\overline{\cal C}} \frac{\rd \om_j}{2 \p \i}
             \overline{F}_{z_j} (\om_j) \biggr]
     \frac{\det_{2m} \Th^{(p)}_{j,k}}
          {\prod_{1 \le j < k \le 2m} (\om_j - \om_k - 2\i)} \epp
\end{multline}
This formula is the main result of our work. It represents the
inhomogeneous density matrix of the integrable spin-1 chain as
a single multiple integral. All dependence on the Trotter number
has been absorbed into the auxiliary functions $G^\pm$ and $S^\pm$.
Therefore the Trotter limit is trivial in this formulation.

Note that it is also easy to perform the homogenous limit. In complete
analogy with the spin-$\2$ case \cite{KMT99b,GKS05} we obtain
\begin{multline} \label{d2homint}
     {D_{[1,m]}}^{\a_1, \dots, \a_m}_{\be_1, \dots, \be_m} (T,h) =
           2^{- m^2 - n_+ (\a) - n_- (\be)} \\
     \biggl[ \prod_{j=1}^p
             \int_{\cal C} \frac{\rd \om_j}{2 \p \i}
             F_{z_j} (\om_j) \biggr]
     \biggl[ \prod_{j = p + 1}^{2m} 
             \int_{\overline{\cal C}} \frac{\rd \om_j}{2 \p \i}
             \overline{F}_{z_j} (\om_j) \biggr]
     \frac{\det_{2m} \Xi^{(p)}_{j,k}}
          {\prod_{1 \le j < k \le 2m} (\om_j - \om_k - 2\i)}
\end{multline}
for the physical density matrix. Here we introduced the notation
\begin{subequations}
\label{matrixxi}
\begin{align}
     \Xi_{j, 2k - 1}^{(p)} & =
	  \frac{\6_\x^{k-1}}{(k-1)!}
          \begin{cases}
             G^+ (\om_j, \x)|_{\x = 0} & j = 1, \dots, p \\
             G^- (\om_j, \x)|_{\x = 0} & j = p + 1, \dots, 2m \epc
          \end{cases} \displaybreak[0] \\[2ex]
     \Xi_{j, 2k}^{(p)} & =
	  \frac{\i \, \6_\x^{k-1}}{(k-1)!}
          \begin{cases}
             S^+ (\om_j, \x)|_{\x = 0} & j = 1, \dots, p \\
             S^- (\om_j, \x)|_{\x = 0} & j = p + 1, \dots, 2m \epp
          \end{cases}
\end{align}
\end{subequations}
\section{One-point functions in factorized form}
\label{sec:onepoint}
In this section we have a closer look at the one-point functions which
are the most elementary correlation functions. Using the general
multiple integral formula (\ref{d2}) we can write the non-zero one-point
functions as
\begin{align} \label{1p2int}
     D^+_+ (\x) & = \frac{\i}{4} \int_\CC \frac{\rd \om_1}{2 \p \i}
                               \int_\CC \frac{\rd \om_2}{2 \p \i}
                  \frac{(\om_1 - \x - \i)(\om_2 - \x - \i)}
                       {\om_1 - \om_2 - 2\i}
                  \biggl|
                  \begin{array}{cc}
                     G^+ (\om_1, \x) & S^+ (\om_1, \x) \\
                     G^+ (\om_2, \x) & S^+ (\om_2, \x)
                  \end{array}
                  \biggr| \epc \notag \\[1ex]
     D^0_0 (\x) & = \frac{\i}{2} \int_\CC \frac{\rd \om_1}{2 \p \i}
                               \int_{\CCq} \frac{\rd \om_2}{2 \p \i}
                  \frac{(\om_1 - \x - \i)(\om_2 - \x + \i)}
                       {\om_1 - \om_2 - 2\i}
                  \biggl|
                  \begin{array}{cc}
                     G^+ (\om_1, \x) & S^+ (\om_1, \x) \\
                     G^- (\om_2, \x) & S^- (\om_2, \x)
                  \end{array}
                  \biggr| \epc \notag \\[1ex]
     D^-_- (\x) & = \frac{\i}{4} \int_{\CCq} \frac{\rd \om_1}{2 \p \i}
                               \int_{\CCq} \frac{\rd \om_2}{2 \p \i}
                  \frac{(\om_1 - \x + \i)(\om_2 - \x + \i)}
                       {\om_1 - \om_2 - 2\i}
                  \biggl|
                  \begin{array}{cc}
                     G^- (\om_1, \x) & S^- (\om_1, \x) \\
                     G^- (\om_2, \x) & S^- (\om_2, \x)
                  \end{array}
                  \biggr| \epp
\end{align}
This is the double integral form of the one-point functions. Judging from
our experience with the spin-$\2$ case \cite{BoKo01,BGKS06} and with the
spin-1 ground state correlation functions \cite{Kitanine01} we expect these
integrals to factorize into sums over products of single integrals.

This is indeed the case. For $n = 0, 1$ we introduce the following
functions represented by single integrals,
\begin{subequations}
\label{defsigmadelta}
\begin{align}
     \s_n (\x) & =
        \int_{\CC^+} \frac{\rd \la}{2 \p \i}
	\frac{\Ff (\la)}{\Bfq (\la)} \la^n G^+ (\la, \x) -
        \int_{\CC^-} \frac{\rd \la}{2 \p \i}
	\frac{\Ffq (\la)}{\Bf (\la)} \la^n G^- (\la, \x) \epc \\[1ex]
     \de_n (\x) & =
        \int_{\CC^+} \frac{\rd \la}{2 \p \i}
	\frac{\Ff (\la)}{\Bfq (\la)} \la^n S^+ (\la, \x) -
        \int_{\CC^-} \frac{\rd \la}{2 \p \i}
	\frac{\Ffq (\la)}{\Bf (\la)} \la^n S^- (\la, \x) \epp
\end{align}
\end{subequations}
Then, using tricks similar to those employed in \cite{BGKS06}, we obtain
the `magnetization'
\begin{equation} \label{magneti}
     D^+_+ (\x) - D^-_- (\x) = \s_0 (\x)
\end{equation}
and the `probability for measuring zero for the $z$-component of the
spin',
\begin{equation} \label{dzz}
     D^0_0 (\x) = \frac{1}{3} - \frac{\i}{3} 
         \left| \begin{array}{cc}
            \s_0 (\x) & \s_1 (\x) - 2 \i \\[2ex]
	    \de_0 (\x) + 1 + \frac{2}{Y(\x)}
	    & \de_1 (\x) + \x \Bigl( 1 + \frac{2}{Y(\x)} \Bigr)
         \end{array} \right| \epc
\end{equation}
in factorized form. They determine all one-point functions because of
the relation
\begin{equation} \label{sumone}
      D^+_+ (\x) + D^0_0 (\x) + D^-_- (\x) = 1 \epp
\end{equation}
Note that $\s_0$ and also the whole determinant in (\ref{dzz}) must
vanish for symmetry reasons if the magnetic field is switched off.

\section{The zero temperature limit at vanishing magnetic field}
All dependence on temperature of the multiple integral formula (\ref{d2})
is hidden in the functions $G^\pm$ and $S^\pm$. We obtain the ground
state result for vanishing magnetic field by replacing these functions
by their corresponding limits which have to be calculated from (\ref{gpm}),
(\ref{spm}).

The temperature enters these equations through the functions $\fb$,
$\fbq$, $\ff$, $\ffq$ and $y$. How do they behave in the limit?
We first look at the nonlinear integral equations (\ref{nlie}). As $T
\rightarrow 0$ for $h = 0$, the driving terms in the equations for
$\fb$ and $\fbq$ both go to minus infinity pointwise. It follows that
$\fb, \fbq \rightarrow 0$ on lines slightly below or slightly above the
real axis. From the equation for $y$ we conclude that $y \rightarrow 1$
close to the real axis. Then by equation (\ref{seconddet}) also
$y(\la \pm \i) \rightarrow 1$ for $\la$ close to the real axis, and,
using (\ref{NLIEf}), we find that $\ff, \ffq \rightarrow 0$. Thus,
\begin{equation} \label{closelim}
     \lim_{T \rightarrow 0+} \lim_{h \rightarrow 0}
        \frac{\Ffq (\la)}{\Bf (\la)} = 1 \epc \qd
     \lim_{T \rightarrow 0+} \lim_{h \rightarrow 0}
        \frac{\Ff (\la)}{\Bfq (\la)} = 1
\end{equation}
for $\la$ slightly below or above the real axis. The behaviour of these
functions close to the lower edge of $\CC^-$ and close to the upper
edge of $\CC^+$ then follows from (\ref{auxid2}),
\begin{equation} \label{farlim}
     \lim_{T \rightarrow 0+} \lim_{h \rightarrow 0}
        \frac{\Ffq (\la - 2\i)}{\Bf (\la - 2\i)} = \2 \epc \qd
     \lim_{T \rightarrow 0+} \lim_{h \rightarrow 0}
        \frac{\Ff (\la + 2\i)}{\Bfq (\la + 2\i)} = \2
\end{equation}
for $\la$ slightly above or below the real axis.

Inserting (\ref{closelim}) and (\ref{farlim}) into (\ref{gpm}) and
(\ref{spm}) and using that $Y (\x) \rightarrow 2$ we obtain a set of
linear integral equations of convolution type that can be solved 
by means of Fourier transformation. Some care is required with the
relative location of the contours, though. Referring to the notation
\begin{subequations}
\begin{align}
     & Z^{++} (\la, \x) = \lim_{\e \rightarrow 0+} \lim_{T \rightarrow 0+}
                          \lim_{h \rightarrow 0} Z^+ (\la + 2\i - \i \e, \x)
                          \epc \\[1ex]
     & Z^{+-} (\la, \x) = \lim_{\e \rightarrow 0+} \lim_{T \rightarrow 0+}
                          \lim_{h \rightarrow 0} Z^+ (\la + \i \e, \x)
                          \epc \\[1ex]
     & Z^{-+} (\la, \x) = \lim_{\e \rightarrow 0+} \lim_{T \rightarrow 0+}
                          \lim_{h \rightarrow 0} Z^- (\la - \i \e, \x)
                          \epc \\[1ex]
     & Z^{--} (\la, \x) = \lim_{\e \rightarrow 0+} \lim_{T \rightarrow 0+}
                          \lim_{h \rightarrow 0} Z^- (\la - 2\i + \i \e, \x)
                          \epc
\end{align}
\end{subequations}
where $Z = G$ or $Z = S$, we obtain the following results
\begin{subequations}
\label{gslimzero}
\begin{align}
     & G^{++} (\la, \x) = G^{--} (\la, \x) = 0 \epc \\[1ex]
     & G^{-+} (\la, \x) = - G^{+-} (\la, \x)
          = \frac{\i \p}{2 \ch \bigl(\frac{\p}{2} (\la - \x) \bigr)}
            \epc \\[1ex]
     & S^{++} (\la, \x) = S^{--} (\la, \x)
          = \frac{\i \p}{\ch \bigl(\frac{\p}{2} (\la - \x) \bigr)}
            \epc \\[1ex]
     & S^{-+} (\la, \x)
          = \frac{\p (\la - \x - 2\i)}
                 {4 \ch \bigl(\frac{\p}{2} (\la - \x) \bigr)} \epc \qd
       S^{+-} (\la, \x)
          = \frac{- \p (\la - \x + 2\i)}
                 {4 \ch \bigl(\frac{\p}{2} (\la - \x) \bigr)} \epp
\end{align}
\end{subequations}

As a first consistency test we may insert these results into our
formulae (\ref{defsigmadelta}) for the one-point functions. We obtain
$\s_0 = 0$ and $\de_0 = - 2$. Then (\ref{magneti}), (\ref{dzz}) and
(\ref{sumone}) imply that $D^+_+ (0) = D^0_0 (0) = D^-_- (0) = 1/3$
as it must be from symmetry considerations. This is, of course, in
agreement with \cite{Kitanine01}.

Still, it is not obvious how to relate, in general, the limit of our
multiple integral to the multiple integral derived there directly
for the ground state at vanishing field. Here we consider only the
case of the one-point functions and defer any further discussion to
future work. We have to calculate the limits of $G^+$ and $S^+$
in the lower strip $\CS^-$ and the limits of $G^-$ and $S^-$ in the
upper strip $\CS^+$ on lines close to the real axis and close to $\pm 2\i$.
These lines must be chosen in such a way that all poles of the kernels in
(\ref{gpm}), (\ref{spm}) are located outside the integration contours.
Keeping this in mind we define
\begin{subequations}
\begin{align}
     & Z_{++} (\la, \x) = \lim_{\e \rightarrow 0+} \lim_{T \rightarrow 0+}
                          \lim_{h \rightarrow 0} Z^+ (\la - \i \e, \x)
                          \epc \\[1ex]
     & Z_{+-} (\la, \x) = \lim_{\e \rightarrow 0+} \lim_{T \rightarrow 0+}
                          \lim_{h \rightarrow 0} Z^+ (\la - 2\i + \i \e, \x)
                          \epc \displaybreak[0] \\[1ex]
     & Z_{-+} (\la, \x) = \lim_{\e \rightarrow 0+} \lim_{T \rightarrow 0+}
                          \lim_{h \rightarrow 0} Z^- (\la + 2\i - \i \e, \x)
                          \epc \\[1ex]
     & Z_{--} (\la, \x) = \lim_{\e \rightarrow 0+} \lim_{T \rightarrow 0+}
                          \lim_{h \rightarrow 0} Z^- (\la + \i \e, \x)
                          \epc
\end{align}
\end{subequations}
for $Z = G$ and $Z = S$. Inserting (\ref{gslimzero}) into (\ref{gpm}),
(\ref{spm}) we obtain
\begin{subequations}
\label{mehrgslim}
\begin{align}
     & G_{-+} (\la, \x) = G_{+-} (\la, \x) = 0 \epc \\[1ex]
     & S_{-+} (\la, \x) = S_{+-} (\la, \x) = 0 \epc \\[1ex]
     & G_{--} (\la, \x) = - G_{++} (\la, \x)
          = \frac{\i \p}{2 \ch \bigl(\frac{\p}{2} (\la - \x) \bigr)}
            \epc \\[1ex]
     & S_{--} (\la, \x)
          = \frac{\p (\la - \x - 2\i)}
                 {4 \ch \bigl(\frac{\p}{2} (\la - \x) \bigr)} \epc \qd
       S_{++} (\la, \x)
          = \frac{- \p (\la - \x + 2\i)}
                 {4 \ch \bigl(\frac{\p}{2} (\la - \x) \bigr)} \epp
\end{align}
\end{subequations}
Inserting (\ref{gslimzero}) and (\ref{mehrgslim}) into (\ref{1p2int}),
in turn, we arrive at
\begin{align} \notag
     D^+_+(0) = & D^-_-(0) = \\ \notag
     = & \frac{\i}{4}
         \int_{- \infty}^\infty \frac{\rd x_1}{\ch({\pi x_1})}
                 \int_{- \infty}^\infty\frac{\rd x_2}{\ch({\pi x_2})}
                 \left[\frac{(x_1+\frac{\i}{2})(x_2-\frac{\i}{2})}
	                    {x_1-x_2+\i0} -
		       \frac{(x_1+\frac{\i}{2})(x_2-\frac{\i}{2})}
	                    {x_1-x_2-\i0} \right] \\
     = & \frac{\i}{4}
         \int_{- \infty}^\infty \frac{\rd x_1}{\ch({\pi x_1} )}
         \int_{- \infty}^\infty \frac{\rd x_2}{\ch({\pi x_2} )}
         \left[\frac{(x_1+\frac{\i}{2})(x_2-\frac{\i}{2})}{x_1-x_2+\i0}
              -\frac{(x_1-\frac{\i}{2})(x_2+\frac{\i}{2})}{x_1-x_2-2\i}
	       \right] \notag \\
                = & \frac{\pi}{2}\int_{- \infty}^\infty \rd x \;
		    \frac{x^2+\frac14}{\ch^2(\pi x)} = \frac13 \epc
\end{align}
to be compared with (4.9) and (4.13) of Kitanine \cite{Kitanine01}.
Similarly, (4.12) of \cite{Kitanine01} for $D^0_0(0)$ is reproduced
as well,
\begin{align}\notag
     D^0_0(0) = & \frac{\i}{2}
                  \int_{- \infty}^\infty \frac{\rd x_1}{\ch({\pi x_1})}
                  \int_{- \infty}^\infty \frac{\rd x_2}{\ch({\pi x_2})}
                  \left[\frac{(x_1+\frac{\i}{2})(x_2+\frac{\i}{2})}
		             {x_1-x_2-\i0}
		       -\frac{(x_1+\frac{\i}{2})(x_2+\frac{\i}{2})}
		             {x_1-x_2+\i0}\right] \\
              = & \frac{\pi}{2} \int_{- \infty}^\infty \rd x \;
	          \frac{\frac12-2x^2}{\ch^2(\pi x)} = \frac13 \epp
\end{align}

\section{Conclusion}
We have managed to represent the inhomogeneous density matrix of the
integrable isotropic spin-$1$ chain as a single multiple integral
(\ref{d2}). Our formula admits of the Trotter limit, the homogeneous
limit and the zero temperature and zero magnetic field limit, where it
reproduces the known values of the one-point functions. The main
difficulty in the derivation of (\ref{d2}) was not in the algebraic part,
which can be treated in a similar way as in the ground state case, but
in the analytic part. For finite temperature we can not work with root
density functions. Instead, the integrals are obtained by replacing sums
over Bethe roots by integrals over closed contours encircling the Bethe
roots. In the spin-$1$ case the Bethe roots for the dominant state of the
quantum transfer matrix come in widely separated pairs, so-called
two-strings. In the Trotter limit they cluster close to $\pm \i$.
Therefore, in order to avoid unwanted extra-terms, we were forced to
introduce closed contours consisting of two separated loops, which brought
about a considerable amount of technical complexity into the derivation as
compared to the spin-$\2$ case \cite{GHS05} (see
appendix~\ref{app:dermult}).

We believe that our result can be generalized to the critical anisotropic
case, as it was done for the ground state at vanishing magnetic field in
\cite{DeMa10}, and to arbitrary higher spins. Of particular interest for
our own research will be the question if the correlation functions of the
integrable higher spin chains factorize. We have obtained a first hint in
this direction: we saw in section \ref{sec:onepoint} that the integrals
for the one-point functions factorize. This is still not what was called
factorization of correlation functions in \cite{BGKS07} and what was
recently proved to hold for the spin-$\2$ XXZ chain, namely, that all
correlation functions (of a suitably regularized model) can be expressed
in terms of a small number of special short-range correlations functions
constituting the `physical part' of the problem (for the physical
part of the XXZ spin-$\2$ correlation functions see \cite{BoGo09}).
Showing this for the higher-spin chains of fusion type as well will be a
challenging project for future research.
\\[1ex]
{\bf Acknowledgment.} We would like to thank T.~Bhattacharyya, H.~Boos,
T.~Deguchi, M.~Jimbo, A.~Kl\"um\-per,~T. Miwa and M.~Takahashi for
stimulating discussions. FG is grateful to Shizuoka University for
hospitality. His work was supported by the DFG under grant number
Go~825/5-1 and by the Volkswagen Foundation. AS gratefully acknowledges
financial support by the DFG under grant number Se 1742/1-2.
JS is supported by a Grant-in-Aid for Scientific Research No.\ 20540370.

\clearpage

{\appendix
\Appendix{Auxiliary functions for spin 1}
\label{app:auxfun}
\noindent
As long as the Trotter number is finite the transfer matrix
eigenvalues $\La^{[1]} (\la)$ and $\La^{[2]} (\la)$ as well as
all the auxiliary functions used in this work can be expressed in
terms of the $Q$-functions (\ref{defq}) and the functions $\Ph_\pm$
defined in (\ref{defphi}). In this appendix we collect the corresponding
formula and also some of the relations between the auxiliary functions.
The presentation largely follows \cite{Suzuki99}.

It is sometimes more convenient to deal with polynomials rather than
with rational functions. For this reason a different normalization of
the elementary $R$-matrix was used in \cite{Suzuki99}. This leads to
differently normalized transfer matrix eigenvalues. In order to simplify
the comparison with \cite{Suzuki99} we define the functions
\begin{subequations}
\begin{align}
     & \La_1 (\la) =
       \phi_- (\la - 3 \i) \phi_+ (\la + 3 \i) \La^{[1]} (\la) \epc \\[1ex]
     & \La_2 (\la) =
       \phi_- (\la - 4 \i) \phi_+ (\la + 2 \i)
       \phi_- (\la - 2 \i) \phi_+ (\la + 4 \i)
       \La^{[2]} (\la) \epp
\end{align}
\end{subequations}
Then, following \cite{Suzuki99}, we introduce
\begin{subequations}
\begin{align}
     & \la_1 (\la) = \re^{- \frac{2h}{T}} \phi_- (\la - 4 \i)
                     \phi_+ (\la - 2 \i) \phi_- (\la - 2 \i) \phi_+ (\la)
                     \frac{q(\la + 3 \i)}{q(\la - \i)} \epc \\[1ex]
     & \la_2 (\la) = \phi_- (\la - 2 \i) \phi_+ (\la) \phi_- (\la)
                     \phi_+ (\la + 2 \i)
                     \frac{q(\la - 3 \i) q(\la + 3 \i)}
                          {q(\la - \i) q(\la + \i)} \epc \\[1ex]
     & \la_3 (\la) = \re^{\frac{2h}{T}} \phi_- (\la) \phi_+ (\la + 2 \i)
                     \phi_- (\la + 2 \i) \phi_+ (\la + 4 \i)
                     \frac{q(\la - 3 \i)}{q(\la + \i)} \epp
\end{align}
\end{subequations}

It follows that
\begin{equation}
     \La_2 (\la) = \la_1 (\la) + \la_2 (\la) + \la_3 (\la) \epp
\end{equation}
The basic auxiliary functions for spin 1 are
\begin{equation}
     \fb (\la) = \frac{\la_1 (\la + \i) + \la_2 (\la + \i)}
                      {\la_3 (\la + \i)} \epc \qd
     \fbq (\la) = \frac{\la_2 (\la - \i) + \la_3 (\la - \i)}
                       {\la_1 (\la - \i)} \epc
\end{equation}
with corresponding capital functions
\begin{equation}
     \Bf (\la) = 1 + \fb (\la) \epc \qd \Bfq (\la) = 1 + \fbq (\la) \epp
\end{equation}

In \cite{Suzuki99} the nonlinear integral equations (\ref{nlie}) were
derived from a set of functional equations satisfied by the functions
$\fb, \fbq, \Bf, \Bfq$ together with
\begin{equation}
     y(\la)  = \frac{\La_2 (\la)}{\phi_- (\la - 4 \i) \phi_+ (\la - 2 \i)
                     \phi_- (\la + 2 \i) \phi_+ (\la + 4 \i)} \epc \qd
     Y(\la) = 1 + y(\la) \epp
\end{equation}
In appendix \ref{app:contnlietonlie} we present an alternative derivation
starting from the integral equations (\ref{LCN}) and combining them with
some of the algebraic relations exposed below.

%For the description of the density matrix we found it useful to
%introduce the functions
%\begin{equation}
%     \ff (\la) = \frac{1}{\fb (\la - 2 \i)} \epc \qd
%     \ffq (\la) = \frac{1}{\fbq (\la + 2 \i)}
%\end{equation}
%and
%\begin{equation}
%     \Ff (\la) = 1 + \ff (\la) \epc \qd \Ffq (\la) = 1 + \ffq (\la) \epp
%\end{equation}

In the derivation of the multiple integral representation for the
density matrix elements we further encounter the functions
\begin{subequations}
\begin{align} \label{defas}
     & \fa (\la) = \frac{d(\la)}{a(\la)}
                   \frac{q(\la + 2 \i)}{q(\la - 2 \i)}
                   \epc \qd \Af (\la) = 1 + \fa (\la) \epc \\[1ex]
     & \faq (\la) = \frac{1}{\fa (\la)} \epc \qd
                    \Afq (\la) = 1 + \faq (\la)
\end{align}
\end{subequations}
familiar from the spin-$\2$ case. We find it also convenient to give a
separate name to the functions with shifted arguments,
\begin{equation} \label{defaII}
     \afII(\la)=\fa(\la+2\i) \epc \quad
     \afIIq(\la)=\faq(\la-2\i) \epp
\end{equation}

The following relations among the functions are needed at several
instances in this work. They follow directly from the above definitions,
\begin{subequations}
\begin{align} \label{auxid1}
     & \frac{\Bf (\la)}{\Ffq (\la)} = \Af (\la + 2 \i) \epc
       \qd \frac{\Bfq (\la)}{\Ff (\la)} = \Afq (\la - 2 \i) \epc \\[1ex]
       \label{auxid2}
     & \Af (\la + \i) \Afq (\la - \i) = 1 + y(\la) \epc \\[1ex]
       \label{auxid3}
     & \frac{\Bf (\la - \i)}{\Af (\la + \i) \Af (\la - \i)} =
       \frac{\La^{[2]} (\la)}{\La^{[1]} (\la + \i) \La^{[1]} (\la - \i)}
       = \frac{1}{1 + y^{-1} (\la)} \epc \\[1ex]
     & \fb(\la-\i) \fbq(\la+\i) = 1+y(\la) = Y(\la) \epc
       \label{bbbarY} \\[1ex]
     & \frac{\Ff (\la) }{\Ffq (\la) } = \frac{\Bfq (\la)}{\Bf (\la)}
       \frac{Y(\la+\i)}{Y(\la-\i)} \fa(\la) \epc \label{FFbar} \\[1ex]
     & \frac{\ffq(\la)}{\afII(\la)}
        = \Bigl(\frac{\Ffq(\la)}{\Bf (\la)} \fa(\la)
	= \Bigr) \frac{\Ffq(\la)}{y(\la+\i)} \epc
          \label{fbarovera2} \\[1ex]
     & \frac{\ff(\la)}{\afIIq(\la)}
        = \Bigl(\frac{\Ff(\la)}{\Bfq(\la)} \faq(\la)
	= \Bigr) \frac{\Ff(\la)}{y(\la-\i)} \epp \label{fovera2b}
\end{align}
\end{subequations}

\Appendix{NLIE with straight contour integrations}
\label{app:contnlietonlie}\noindent
In this appendix we will show the steps that are necessary for transforming
(\ref{LCN}) into (\ref{nlie}). We also present subsidiary equations
which can be used for the numerical calculation of some of the auxiliary
functions on lines away from the real axis.

First note that numerical calculations with fixed Trotter number $N$
suggest that
\begin{align}
     & |\fbq(\la)|,\ |\ff(\la)|  \ll 1 \quad 
       \text{for } \Im \la = \epsilon, & 
     & |\fb(\la)|,\ |\ffq(\la)|  \ll 1 \quad
       \text{for }  \Im \la = -\epsilon \epc \\
     & |\fbq(\la)|,\ |\ff(\la)|  \gg 1 \quad
       \text{for }  \Im \la = 2-\epsilon, & 
     & |\fb(\la)|,\ |\ffq(\la)|  \gg 1 \quad
       \text{for }  \Im \la = -2+\epsilon \notag
\end{align}
in the low temperature regime. Therefore we rewrite, for example, 
\begin{multline}
     \int_{\CC^+} \frac{\rd \m}{2 \p \i} \jg2(\la-\mu) \ln \Ff(\mu) \\ =
     \int_{-\infty+\i\epsilon}^{\infty+\i\epsilon} \frac{\rd \m}{2 \p \i}
        \jg2(\la-\mu) \ln \Ff(\mu)
     - \int_{-\infty-\i\epsilon}^{\infty-\i\epsilon} \frac{\rd \m}{2 \p \i}
        \jg2(\la-\mu-2\i) \ln \frac{\Bf(\mu)}{\fb(\mu)}
\end{multline}
for $\la$ located inside a narrow strip ${\mathcal S}_0$ including
the real axis. To emphasize the relative location of $\la$ and $\mu$,
we write the last integral as
\begin{equation}
     - \int_{\Im \la>\Im \mu} \frac{\rd \m}{2 \p \i} \jg2(\la-\mu-2\i)
       \ln \frac{\Bf(\mu)}{\fb(\mu)} \epp
\end{equation}

%In the following we shall need further relations among the auxiliary
%functions,
%\begin{subequations}
%\begin{align}
%\end{align}
%\end{subequations}
We keep our assumption that $\la \in {\mathcal S}_0$ for a while. Thanks
to (\ref{fbarovera2}), (\ref{fovera2b}) and a similar transformation
applied to the integrands, (\ref{LCN}) is represented as
\begin{subequations}
\begin{align}
     & \ln y(\la-\i) = -d_f(\la) +\ln \Bfq (\la) \notag \\
     & \phantom{ccc} + \int_{\Im \la>\Im \mu} \frac{\rd \m}{2 \p \i}
       \jg2(\la-\mu-2\i) \ln \frac{\Bf (\mu)}{\fb(\mu)}
       - \int_{\Im \la<\Im \mu} \frac{\rd \m}{2 \p \i}
       \jg2(\la-\mu+2\i) \ln \frac{\Bfq (\mu)}{\fbq (\mu)} \notag \\
     & \phantom{ccc} + \int_{-\infty}^{\infty} \frac{\rd \m}{2 \p \i}
       \jg2(\la-\mu+\i\epsilon ) \ln  \Ffq(\mu-\i\epsilon )
       - \int_{-\infty}^{\infty} \frac{\rd \m}{2 \p \i}
       \jg2(\la-\mu-\i\epsilon) \ln \Ff(\mu+\i\epsilon)
       \epc \label{lny1} \\[1ex]
     & \ln y(\la+\i) = d_f(\la) + \ln \Bf (\la) \notag \\
     & \phantom{ccc} - \int_{\Im \la>\Im \mu} \frac{\rd \m}{2 \p \i}
       \jg2(\la-\mu-2\i) \ln \frac{\Bf (\mu)}{\fb(\mu)}
       + \int_{\Im \la<\Im \mu} \frac{\rd \m}{2 \p \i}
       \jg2(\la-\mu+2\i) \ln \frac{\Bfq (\mu)}{\fbq (\mu)} \notag \\
     & \phantom{ccc} - \int_{-\infty}^{\infty} \frac{\rd \m}{2 \p \i}
       \jg2(\la-\mu+\i\epsilon ) \ln  \Ffq(\mu-\i\epsilon )
       + \int_{-\infty}^{\infty} \frac{\rd \m}{2 \p \i}
       \jg2(\la-\mu-\i\epsilon) \ln  \Ff(\mu+\i\epsilon) \epp \label{lny2}
\end{align}
\end{subequations}
The integrands in the last two terms in (\ref{lny1}) and (\ref{lny2})
become proportional to the logarithm of $\Ff (\la) /\Ffq (\la)$
in $\epsilon \rightarrow 0$ limit. Since such ratio does not appear in
(\ref{nlie}), we would like to replace it using (\ref{FFbar}). For this
purpose, we first note a contour integral representation for
$\ln \fa(\la)$,
\begin{equation}
     \ln \fa(\la) = - d_f(\la)
        - \int_{\CC^+} \frac{\rd \m}{2 \p \i} \jg2(\la-\mu) \ln \Ff(\mu)
	- \int_{\CC^-} \frac{\rd \m}{2 \p \i} \jg2(\la-\mu) \ln \Ffq(\mu)
	  \epp
\end{equation}
Again we rewrite this using integration on straight lines and substitute
the result into (\ref{FFbar}). It is then immediately clear that 
\begin{align} \label{FFbarSCN} 
     \ln & \frac{\Ff (\la)}{\Ffq (\la)} =
        - d_f(\la) + \ln \frac{\Bfq (\la)} {\Bf(\la)}
	+ \ln \frac{Y(\la+\i)} {Y(\la-\i)}  \notag \\
     & + \int_{\Im \la>\Im \mu} \frac{\rd \m}{2 \p \i}
        \jg2(\la-\mu-2\i) \ln \frac{\Bf (\mu)}{\fb(\mu)}
       - \int_{\Im \la<\Im \mu} \frac{\rd \m}{2 \p \i}
        \jg2(\la-\mu+2\i) \ln \frac{\Bfq (\mu)}{\fbq (\mu)} \notag \\
     & + \int_{-\infty}^{\infty} \frac{\rd \m}{2 \p \i}
        \jg2(\la-\mu+\i\epsilon ) \ln \Ffq(\mu-\i\epsilon )
       - \int_{-\infty}^{\infty} \frac{\rd \m}{2 \p \i}
        \jg2(\la-\mu-\i\epsilon) \ln  \Ff(\mu+\i\epsilon) \epp
\end{align}

To proceed further, it is convenient to consider equations in Fourier
space. For a smooth function $f(\la)$ we define 
\begin{equation}
     \hat{f}(k) =
        \int_{-\infty}^{\infty} \frac{\rd \la}{2\pi}
	{\rm e}^{\i k\la} f(\la) \epc \qd
%& f(\la) & = \int_{-\infty}^{\infty} \rd k\,
%{\rm e}^{- \i k\la} \hat{f}(k) \epc \\
     \hat{{\rm dl}}f(k) = \int_{-\infty}^{\infty}
	\frac{\rd \la}{2\pi} {\rm e}^{\i k\la}
        \Bigl(\frac{d}{d\la} \ln f(\la) \Bigr) \epp
\end{equation}
We also introduce shifted functions
\begin{subequations}
\begin{align}
     \fb_{\epsilon}(\la) & = \fb(\la-\i\epsilon) \epc &
     \fbq_{\epsilon}(\la) & = \fbq(\la+\i\epsilon) \epc \\
     \ff_{\epsilon}(\la) & = \ff(\la+\i\epsilon) \epc &
     \ffq_{\epsilon}(\la)&=\ffq(\la-\i\epsilon) \epc
\end{align}
\end{subequations}
and similarly for the capital functions.

First we take the Fourier transformation of (\ref{bbbarY}) for
$\lambda$ real. This leads to  a direct relation  between 
$\hat{{\rm dl}} \fb_{\epsilon} (k)$ and $\hat{{\rm dl}} \fbq_{\epsilon}
(k)$,
\begin{equation} \label{bbarFT}
     {\rm e}^{-\epsilon k} \hat{{\rm dl}} \fbq_{\epsilon} (k) =
        - {\rm e}^{-(2-\epsilon)k} \hat{{\rm dl}} \fb_{\epsilon} (k)
	+ { \rm e}^{-k} \hat{{\rm dl}} Y(k) - \i N {\rm e}^{-k} \sinh uk
	\epp
\end{equation}
Similarly, take the Fourier transformation of (\ref{FFbarSCN}) and delete
$\hat{{\rm dl}} \fbq_{\epsilon} (k)$ by means of (\ref{bbarFT}). Then
\begin{align}
     & {\rm e}^{-\epsilon k} \hat{{\rm dl}} \Ff_{\epsilon} (k)
       - {\rm e}^{\epsilon k} \hat{{\rm dl}} \Ffq_{\epsilon} (k)
       = \frac{1}{1+K_2(k)} \Bigl(
          - \hat{{\rm dl}} \Delta_f(k)+{\rm e}^{\epsilon k} (K_{yb}(k) -1)
	    \hat{{\rm dl}} \Bf_{\epsilon} (k) \notag \\
     & \phantom{cccc} - {\rm e}^{-\epsilon k} (K_{y{\bar b}}(k) -1)
       \hat{{\rm dl}} \Bfq_{\epsilon}(k) - {\rm e}^{\epsilon k}
       (K_{yb}(k)+{\rm e}^{-2k}  K_{y{\bar b}}(k) ) \hat{{\rm dl}}
       \fb_{\epsilon} (k) \notag \\
     & \phantom{cccc} + ({\rm e}^{-k}  K_{y{\bar b}}(k)
                         + {\rm e}^{k} - {\rm e}^{-k} )
			 \hat{{\rm dl}} Y(k) \Bigr) \epc
			 \label{FFbarFT} \\[2ex]
     & \hat{{\rm dl}} \Delta_f(k) =
       \i N {\rm  e}^{-2|k|-k} \sinh uk \epc \notag \\[1ex]
%
%\hat{{\rm dl}} d_f (k) +N i {\rm e}^{-k} K_{y{\bar b}}(k)  \sinh uk
%\nonumber \\
%
%     \phantom{cccc} & K_{yb}(k) =
%     \begin{cases} {\rm e}^{-4k} & k>0 \\ 1 & k<0 \end{cases} \epc \qquad
%     K_{y\bar{b}}(k) =
%     \begin{cases} 1 & k>0 \\ {\rm e}^{4k} & k<0 \end{cases} \epc \qquad 
     \phantom{cccc} & K_2(k)={\rm e}^{-2|k|} \epc \qd
                      K_{yb}(k) = \re^{- 2 (|k| + k)} \epc \qd
                      K_{y\bar{b}}(k) = \re^{- 2 (|k| - k)} \epp \notag
\end{align}

Finally take the Fourier transformation of the logarithmic derivatives of
both sides of (\ref{lny1}) and (\ref{lny2}). Note that $\hat{{\rm dl}}
\Ff_{\epsilon}(k)$ and $\hat{{\rm dl}} \Ffq_{\epsilon} (k)$ only appear
in the combination ${\rm e}^{-\epsilon k} \hat{{\rm dl}}
\Ff _{\epsilon}(k) - {\rm e}^{\epsilon k} \hat{{\rm dl}}
\Ffq_{\epsilon} (k)$. Therefore, by substituting (\ref{bbarFT}) and
(\ref {FFbarFT}), one obtains two equations containing $\hat{{\rm dl}} 
\fb_{\epsilon} $,  $\hat{{\rm dl}} y  $, $\hat{{\rm dl}} \Bf_{\epsilon},
\hat{{\rm dl}} \Bfq_{\epsilon} $ and $\hat{{\rm dl}} Y$. They can be
solved for  $\hat{{\rm dl}} \fb_{\epsilon}$ and  $\hat{{\rm dl}} y$ in
terms of $\hat{{\rm dl}} \Bf_{\epsilon}, \hat{{\rm dl}} \Bfq_{\epsilon}$
and $\hat{{\rm dl}} Y$, yielding% (for $k>0$)
\begin{subequations}
\begin{align}
     \hat{{\rm dl}} \fb_{\epsilon} (k) &
        = - \i N \frac{{\rm e}^{-\epsilon k} \sinh u k }{2 \cosh k}  +
            \frac{ \hat{{\rm dl}} \Bf _{\epsilon}(k)}
	         {1+{\rm e}^{2|k|}}
          - \frac{\hat{{\rm dl}} \Bfq_{\epsilon} (k)}
	         {1+{\rm e}^{2|k|}}{\rm e}^{2(1 - \epsilon) k}
          + \frac{ \hat{{\rm dl}}Y (k)}{2\cosh k} {\rm e}^{-\epsilon k}
	  \epc \label{bFT} \\[1ex]
     \hat{{\rm dl}} y (k) & =
        \frac{{\rm e}^{\epsilon k} \hat{{\rm dl}} \Bf _{\epsilon}(k)
	    + {\rm e}^{-\epsilon k}\hat{{\rm dl}} \Bfq_{\epsilon} (k)}
	     {2\cosh k} \epp
\end{align}
\end{subequations}
If  $\hat{{\rm dl}} \fb_{\epsilon} (k)$ is eliminated from (\ref{bFT})
by means of (\ref{bbarFT}), an equation for $\hat{{\rm dl}}
\fbq_{\epsilon} (k)$ is obtained,
\begin{equation}
     \hat{{\rm dl}} \fbq_{\epsilon} (k) =
        - \i N \frac{{\rm e}^{\epsilon k} \sinh u k }{2 \cosh k}
	+ \frac{ \hat{{\rm dl}} \Bfq_{\epsilon} (k)}{1+{\rm e}^{2|k|}}
        - \frac{ \hat{{\rm dl}} \Bf_{\epsilon} (k)}
	       {1+{\rm e}^{2|k|}} {\rm e}^{-2(1-\epsilon) k}
        + \frac{ \hat{{\rm dl}}Y (k)}{2\cosh k} {\rm e}^{\epsilon k} \epp
\end{equation}
%The similar transformation for $k<0$ is possible.
Applying the inverse Fourier transformation and integrating once, we
successfully recover the NLIE (\ref{nlie}) with straight integration
contours.

To evaluate physical quantities beyond $\Lambda^{[2]}$, we also need NLIE
(defined with straight integration contours) for $\ff_{\epsilon}$ and
$\ffq_{\epsilon}$. This can be understood as follows. The eigenvalues of
physical quantities are parameterized by BAE roots. Thus, they can be
naturally represented by loop integrals involving $\Bf(\la), \Bfq(\la),
\Ff(\la)$ or  $\Ffq(\la)$. We consider, for example, 
\begin{equation}
     {\rm I} = \int_{\CC^-} \rd \m\, \frac{P(\la,\mu)}{\Bf(\mu)} \epc
\end{equation}
where $P(\la,\mu)$ is some function.
This integral can be represented as 
\begin{equation}
     {\rm I} = - \int_{-\infty}^{\infty} \rd \m\,
                 \frac{P(\la,\mu-\i\epsilon)}{1+\fb_{\epsilon}(\mu)}
               + \int_{-\infty}^{\infty} \rd \m\,
	         \frac{P(\la,\mu-2\i+\i\epsilon)}
		      {1+\frac{1}{\ff_{\epsilon}(\mu)}} \epp
\end{equation}
We therefore need to evaluate $ \ff_{\epsilon}(\la)$ and
$\ffq_{\epsilon}(\la)$, when we adopt straight lines near the real axis
as integration contours.

Indeed, it is not difficult to derive the following expressions for
$\ff_{\epsilon}(\la)$ and $\ffq_{\epsilon}(\la)$,
\begin{subequations}
\label{NLIEf}
\begin{align}
     \ln \ff_{\epsilon} (\la) = &
        \D_{\overline{b}}(\la) +
	\int_{-\infty}^{\infty} \rd \m\, \kmat_{\bar{b}b}(\la-\mu)
	\ln \Bf_{\epsilon}(\mu) \notag \\ & +
	\int_{-\infty}^{\infty} \rd \m\, \kmat_{\bar{b}\bar{b}}(\la-\mu)
	\ln \Bfq_{\epsilon}(\mu) +
	\int_{-\infty}^{\infty} \rd \m\,
        \kmat_{fy}(\la-\mu) \ln Y_-(\mu) \epc
	\label{NLIEf1} \\[1ex]
     \ln \ffq_{\epsilon}(\la) = &
        \D_{b}(\la) +
	\int_{-\infty}^{\infty} \rd \m\, \kmat_{bb}(\la-\mu)
	\ln \Bf_{\epsilon}(\mu) \notag \\ & +
	\int_{-\infty}^{\infty} \rd \m\, \kmat_{b\bar{b}}(\la-\mu)
	\ln \Bfq_{\epsilon}(\mu) +
	\int_{-\infty}^{\infty} \rd \m\,  \kmat_{\bar{f}y}(\la-\mu)
	\ln Y_+(\mu) \epp
	\label{NLIEf2}
\end{align}
\end{subequations}
The integration kernel $\kmat_{ab}$ is the corresponding component in
(\ref{ksym}), except for $ \kmat_{fy}$ and $\kmat_{\bar{f}y}$,
defined explicitly by $\kmat_{fy}(\la)= - {\cal K}(\la - \i (1-\epsilon))$
and $\kmat_{\bar{f}y}(\la)=-{\cal K}(\la + \i (1-\epsilon))$. 

The functions $Y_{\pm}(\la)$ denote shifted $Y$-functions, 
$Y_{\pm}(\la) = Y(\la\pm \i)$. Unfortunately, they can not be determined
from (\ref{nlie}), because of the singularity of the kernel function.
We thus need subsidiary equations,
\begin{subequations}
\label{seconddet}
\begin{align}
     \ln y_+(\la) = \ln \Bf_{\epsilon}(\la+\i\epsilon) +
        \int_{-\infty}^{\infty} \rd & \m\, \kmat_{yb}(\la-\mu+\i)
	\ln \Bf_{\epsilon} (\mu) \notag \\
        & + \int_{-\infty}^{\infty} \rd \m\, \kmat_{y\bar{b}}(\la-\mu+\i)
	\ln \Bfq_{\epsilon} (\mu) \epc \\[1ex]
     \ln y_-(\la) = \ln \Bfq_{\epsilon}(\la-\i\epsilon) +
        \int_{-\infty}^{\infty} \rd & \m\, \kmat_{yb}(\la-\mu-\i)
	\ln \Bf_{\epsilon} (\mu) \notag \\
        & + \int_{-\infty}^{\infty} \rd \m\, \kmat_{y\bar{b}}(\la-\mu-\i)
	\ln \Bfq_{\epsilon} (\mu) \epp
\end{align}
\end{subequations}
The functions $\fb(\la)$ and $\fbq(\la)$ are analytic in a narrow strip
including the real axis. For this reason we can use (\ref {nlie}) to
estimate the first terms in the rhs of (\ref{seconddet}). Thus,
(\ref{NLIEf}) and  (\ref{seconddet}), together with (\ref{nlie}),
fix $\ff(\la)$ and $\ffq(\la)$ through integrals defined on straight
contours.
\enlargethispage{-2ex}

%==========================================================
\Appendix{Derivation of the multiple integral representation}
\label{app:dermult}
\noindent
In this appendix we derive the multiple integral representation of section
\ref{sec:mult}. Our strategy is to use as much as possible the results
obtained in \cite{GHS05} for the spin-1/2 case.
\subsection{Results for spin-1/2 auxiliary space}
\subsubsection{Spin projection conserving basis}
The monodromy matrix $T^{[1]}$ preserves the pseudo spin projection
\begin{equation}
     \h^z = \sum_{j=1}^N (-1)^j S_j^z \epc \qd
        [ T_a^{[1]} (\la), \tst{\2} \s_a^z + \h^z] = 0 \epp
\end{equation}
It follows that
\begin{equation}
     {T^{[1]}}^{\a_1}_{\be_1} (\z_1) \dots
        {T^{[1]}}^{\a_n}_{\be_n} (\z_n) \h^z = 
        \Bigl( \h^z + \tst{\2} \sum_{j=1}^n (\a_j - \be_j) \Bigr)
        {T^{[1]}}^{\a_1}_{\be_1} (\z_1) \dots
        {T^{[1]}}^{\a_n}_{\be_n} (\z_n) \epp
\end{equation}
Since the dominant state $|\Ps_0\> = B(\la_1) \dots B(\la_N) |0\>$ has
pseudo spin projection zero, $\h^z |\Ps_0\> = 0$, we conclude that the
matrix elements $\<\Ps_0|{T^{[1]}}^{\a_1}_{\be_1} (\z_1) \dots
{T^{[1]}}^{\a_n}_{\be_n} (\z_n)|\Ps_0\>$ all vanish, unless $\sum_{j=1}^n
(\a_j - \be_j) = 0$.

This means that we must have the same number of plus signs in the sequences
$(\a_j)$ and $(\be_k)$ of upper and lower indices. Let us introduce
a basis on the space of local operators which is adapted to this fact.
It is convenient to label the states in this basis by the positions
of the plus signs in $(\a_j)$ and minus signs in $(\be_k)$. For
$\xv = (x_1, \dots, x_n)$ with $x_j \in {\mathbb Z}_n = \{1, \dots, n\}$
and $\{x_1, \dots, x_p\}$, $\{x_{p+1}, \dots, x_n\}$ two sets of mutually
distinct numbers, let
\begin{equation}
     b_p (\xv) = \s_{x_n}^- \dots \s_{x_{p+1}}^- \s_1^+ \dots \s_n^+
                 \s_{x_p}^- \dots \s_{x_1}^- \epp 
\end{equation}
Then
\begin{equation}
     b_p (\xv) = {e_1}^{\a_1}_{\be_1} \dots {e_n}^{\a_n}_{\be_n}
        \qd \text{with} \qd
        \begin{array}{l}
        \a_j = \begin{cases}
                  + & \text{if $j \in \{x_1, \dots, x_p\}$}\\
                  - & \text{else}
               \end{cases} \\[3ex]
        \be_j = \begin{cases}
                   + & \text{if $j \notin \{x_{p+1}, \dots, x_n\}$}\\
                   - & \text{else.}
                \end{cases}
        \end{array}
\end{equation}
Clearly
\begin{equation}
     B_n = \bigl\{ b_p (\xv) \big| n \ge x_1 > \dots > x_p \ge 1
                   \le x_{p+1} < \dots < x_n \le n; p = 0, \dots, n \bigr\}
\end{equation}
is a basis of the $\h^z = 0$ subspace of the space of local operators
acting on $\bigl({\mathbb C}^2\bigr)^{\otimes n}$.
\subsubsection{Combinatorial formula for density matrix at finite Trotter
number}
Referring to the notation of the previous subsection we now fix an even
$n = 2m$ and a vector $\xv$ that specifies a basis element in $B_{2m}$.
We further define $\z = (\z_1, \dots, \z_{2m})$ and
\begin{equation}
     D^{[1]} (\xv| \z) = \frac{\<\Ps_0| \tr\{ T^{[1]}(\z_1) \otimes \dots
                                \otimes T^{[1]}(\z_{2m}) b_p^t (\xv) \}
				|\Ps_0\>}
                              {\<\Ps_0|\Ps_0\> \La^{[1]} (\z_1) \dots
                               \La^{[1]} (\z_{2m})} \epp
\end{equation}
Density matrix elements of this form were considered in \cite{GHS05},
where a multiple integral representation for the spin-1/2 XXZ chain
at finite temperature was derived. Most of that calculation, up to the
very last step, was purely algebraic and entirely based on the
commutation relations between the elements of the monodromy matrix.
This means it only depended on the structure of the $R$-matrix and,
hence, can be taken over to the present case.

For this purpose let us first of all recall some of the notation of
\cite{GHS05}, but in a form already adapted to the rational limit.
Let
\begin{subequations}
\begin{align}
     \F_j (\la) & = \prod_{k=1}^{x_j - 1} (\la - \z_k - 2\i)
                   \prod_{k = x_j + 1}^{2m} (\la - \z_k) \epc \qd
                   j = 1, \dots, p \epc \\
     \Fq_j (\la) & = \prod_{k=1}^{x_j - 1} (\la - \z_k + 2\i)
                   \prod_{k = x_j + 1}^{2m} (\la - \z_k) \epc \qd
                   j = p + 1, \dots, 2m \epc
\end{align}
\end{subequations}
and define a set of functions $w_j (\z)$, $j = 1, \dots, N$, as the
solutions of the linear system
\begin{equation} \label{defw}
     \fa' (\la_j) w_j (\z) = \re(\z - \la_j) \fa(\z) - \re(\la_j - \z)
                             + \sum_{k=1}^N K(\la_j - \la_k) w_k (\z) \epc
\end{equation}
where $\fa$ is the auxiliary function defined in (\ref{defas}).

Then, from equation (63) of \cite{GHS05}, we have the following
combinatorial expression
\begin{multline} \label{ghs1}
     D^{[1]} (\xv| \z) =
     \sum_{(\{\e^+\},\{\e^-\}) \in p_2 ({\mathbb Z}_{2m})} \qd
     \sum_{\ell_{\e_1^+}, \dots, \ell_{\e_{2m-n}^+} = 1}^N \qd
     \sum_{\substack{(\{\de^+\}, \{\de^-\}) \in p_2 ({\mathbb Z}_{2m}) \\
                     \card \{\de^-\} = n}} \sign(PQ) \\[1ex]
     \Biggl[ \prod_{j = 1}^{2m} \frac{1}{1 + \fa(\z_j)} \Biggr]
     \sum_{R \in {\mathfrak S}^n}
     \frac{\sign(R) \det\bigl[ - w_{\ell_{\e_j^+}} (\z_{\de_k^+})\bigr]}
	  {\prod_{1 \le j < k \le 2m} (\z_k - \z_j)(\om_j - \om_k - 2\i)}
           \biggr|_{\om_{\e_j^+} = \la_{\ell_{\e_j^+}}, \:
	           \om_{\e_j^-} = \z_{\de_{Rj}^-}} \\[1ex]
     \Biggl[ \prod_{\substack{j = 1\\ \e_j^+ \le p}}^{2m-n}
	     \F_{\e_j^+} (\la_{\ell_{\e_j^+}}) \Biggr]
     \mspace{-3.mu}
     \Biggl[ \prod_{\substack{j = 1\\ \e_j^+ > p}}^{2m-n}
             - \Fq_{\e_j^+} (\la_{\ell_{\e_j^+}}) \Biggr]
     \mspace{-3.mu}
     \Biggl[ \prod_{\substack{j = 1\\ \e_j^- \le p}}^n
             \F_{\e_j^-} (\z_{\de_{Rj}^-}) \Biggr]
     \mspace{-3.mu}
     \Biggl[ \prod_{\substack{j = 1\\ \e_j^- > p}}^n
             \fa (\z_{\de_{Rj}^-})
             \Fq_{\e_j^-} (\z_{\de_{Rj}^-}) \Biggr].
\end{multline}
For the sums we have adopted the notation from \cite{GHS05}.
${\mathbb Z}_{2m} = \{1, \dots, 2m\}$, and $p_2 ({\mathbb Z}_{2m})$ is
the set of all partitions of ${\mathbb Z}_{2m}$ into ordered pairs of
disjoint subsets. E.g.\ the first sum is over all pairs $(\{\e^+\},
\{\e^-\})$ with $\{\e^+\} \cup \{\e^-\} = {\mathbb Z}_{2m}$ and
$\{\e^+\} \cap \{\e^-\} = \emptyset$. Moreover, $n = \card\{\e^-\}$
by definition. We enumerate the elements in the sets $\{\e^\pm\}$ and
$\{\de^\pm\}$ in such a way that $\e_j^\pm < \e_k^\pm$  and $\de_j^\pm
< \de_k^\pm$ if $j < k$. Then for every $(\{\e^+\},\{ \e^-\})$ and
$(\{\de^+\}, \{\de^-\})$ the permutations $P, Q \in {\mathfrak S}^{2m}$
under the sum are fixed by
\begin{equation}
     Pj = \begin{cases}
             \e_j^- & j = 1, \dots, n \\
             \e_{j-n}^+ & j = n+1, \dots, 2m \epc
          \end{cases} \qd
     Qj = \begin{cases}
             \de_j^- & j = 1, \dots, n \\
             \de_{j-n}^+ & j = n+1, \dots, 2m \epp
          \end{cases}
\end{equation}
Note that the Bethe equations $\fa (\la_j) = - 1$ were used in the
derivation of (\ref{ghs1}).
\subsection{Fusion for density matrix elements}
\subsubsection{The narrow contour}
\enlargethispage{-2ex}
We shall employ equation (\ref{ghs1}) in the derivation of a multiple
integral representation for the density matrix elements of the spin-1
chain. We begin by fixing real inhomogeneity parameters $\x_j$, $j = 1,
\dots, m$, and $\e > 0$. For $j = 1, \dots, m$ we choose $\de_j \in
\{1 + \e, - 1 - \e\}$ arbitrarily and define
\begin{equation}
     \z_{2j - 1} = \x_j + \i \de_j \epc \qd
     \z_{2j} = \x_j - \i \de_j \epp
\end{equation}
Using the fusion formulae (\ref{spin1mono}) and (\ref{monolr}) we can
express the inhomogeneous density matrix (\ref{d2inhom}) of the spin-1
chain as
\begin{multline}
     D^{[2]} (\x) = \lim_{\e \rightarrow 0+}
        \frac{S^{\otimes m} \<\Ps_0| T^{[1]} (\z_1) \otimes \dots \otimes
              T^{[1]} (\z_{2m}) |\Ps_0\> (S^t)^{\otimes m}}
             {\<\Ps_0|\Ps_0\> \La^{[2]} (\x_1) \dots \La^{[2]} (\x_m)}
             \\[2ex]
        = \sum_{\xv \in B_{2m}} S^{\otimes m} b_p (\xv) (S^t)^{\otimes m}
          \lim_{\e \rightarrow 0+}
          \frac{\<\Ps_0| \tr\{T^{[1]} (\z_1) \otimes \dots \otimes
              T^{[1]} (\z_{2m}) b_p^t(\xv)\} |\Ps_0\>}
             {\<\Ps_0|\Ps_0\> \La^{[2]} (\x_1) \dots \La^{[2]} (\x_m)}
             \displaybreak[0] \\[1ex]
        = \sum_{\xv \in B_{2m}} S^{\otimes m} b_p (\xv) (S^t)^{\otimes m}
          \biggl[ \prod_{j=1}^m
             \frac{\La^{[1]} (\x_j - \i) \La^{[1]} (\x_j + \i)}
                  {\La^{[2]} (\x_j)} \biggr]
          \lim_{\e \rightarrow 0+} D^{[1]} (\xv| \z) \epp
\end{multline}
Let us denote the coefficient under the sum
\begin{equation}
     D^{[2]} (\xv| \x) =
          \biggl[ \prod_{j=1}^m
             \frac{\La^{[1]} (\x_j - \i) \La^{[1]} (\x_j + \i)}
                  {\La^{[2]} (\x_j)} \biggr]
          \lim_{\e \rightarrow 0+} D^{[1]} (\xv| \z) \epp
\end{equation}
Inserting equations (\ref{ghs1}) and (\ref{auxid3}) on the right
hand side we obtain
\begin{multline} \label{dens2x1}
     D^{[2]} (\xv| \z) = \lim_{\e \rightarrow 0+}
     \sum_{(\{\e^+\},\{\e^-\}) \in p_2 ({\mathbb Z}_{2m})} \qd
     \sum_{\ell_{\e_1^+}, \dots, \ell_{\e_{2m-n}^+} = 1}^N \qd
     \sum_{\substack{(\{\de^+\}, \{\de^-\}) \in p_2 ({\mathbb Z}_{2m}) \\
                     \card \{\de^-\} = n}} \sign(PQ) \\[1ex]
     \Biggl[ \prod_{j = 1}^m \frac{1}{\Bf(\x_j - \i)} \Biggr]
     \sum_{R \in {\mathfrak S}^n}
     \frac{\sign(R) \det\bigl[ - w_{\ell_{\e_j^+}} (\z_{\de_k^+})\bigr]}
	  {\prod_{1 \le j < k \le 2m} (\z_k - \z_j)(\om_j - \om_k - 2\i)}
           \biggr|_{\om_{\e_j^+} = \la_{\ell_{\e_j^+}}, \:
	           \om_{\e_j^-} = \z_{\de_{Rj}^-}} \\[1ex]
     \Biggl[ \prod_{\substack{j = 1\\ \e_j^+ \le p}}^{2m-n}
	     \F_{\e_j^+} (\la_{\ell_{\e_j^+}}) \Biggr]
     \Biggl[ \prod_{\substack{j = 1\\ \e_j^+ > p}}^{2m-n}
             - \Fq_{\e_j^+} (\la_{\ell_{\e_j^+}}) \Biggr]
     \Biggl[ \prod_{\substack{j = 1\\ \e_j^- \le p}}^n
             \F_{\e_j^-} (\z_{\de_{Rj}^-}) \Biggr]
     \Biggl[ \prod_{\substack{j = 1\\ \e_j^- > p}}^n
             \fa (\z_{\de_{Rj}^-})
             \Fq_{\e_j^-} (\z_{\de_{Rj}^-}) \Biggr] \epp
\end{multline}
Here the limit in the explicit term is easy to calculate
\begin{multline} \label{zetaprod}
     \lim_{\e \rightarrow 0+}
        \prod_{1 \le j < k \le 2m} (\z_k - \z_j) =
     \lim_{\e \rightarrow 0+}
        \biggl[ \prod_{j=1}^m (\z_{2j} - \z_{2j-1}) \biggr] \\[1ex] \times
        \prod_{1 \le j < k \le m} (\z_{2k} - \z_{2j-1}) (\z_{2k} - \z_{2j})
        (\z_{2k-1} - \z_{2j-1}) (\z_{2k-1} - \z_{2j}) \\[1ex] =
        \biggl[ \prod_{j=1}^m - 2 \i \sign \de_j \biggr]
        \prod_{1 \le j < k \le m} (\x_k - \x_j)^2 [(\x_k - \x_j)^2 + 4]
        \epp
\end{multline}

%=======================================================================%
\begin{figure}
\begin{center}
\includegraphics[height=16em]{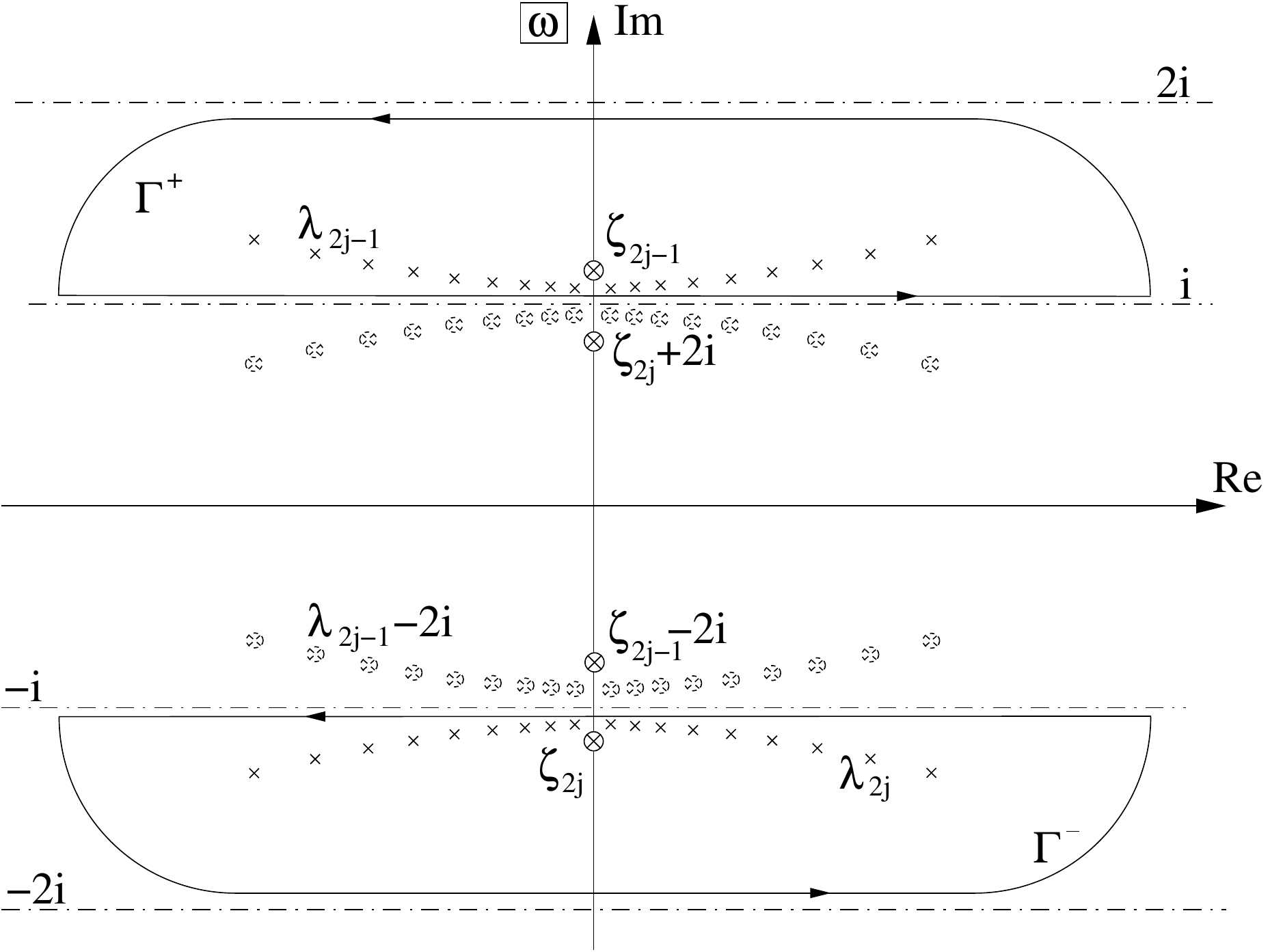} 
\caption{\label{fig:spin1Gamma} The contours $\G^\pm$ only contain Bethe
roots $\{\la_k\}=\{\la_{2j}\}\cup\{\la_{2j-1}\}$ and the corresponding
inhomogeneities $\{\z_k\}=\{\z_{2j}\}\cup\{\z_{2j-1}\}$. The Bethe roots
and inhomogeneities shifted by the amount of $2\i$ are located outside.}
\end{center}
\end{figure}
%=======================================================================%

The combinatorial sum can be converted into a multiple integral by the
same token as in equation (65) of \cite{GHS05}. We introduce a function
\begin{equation} \label{defchi}
     \chi (\la, \z) = \re (\z - \la) \fa (\z) - \re (\la - \z)
                      + \sum_{k=1}^N K(\la - \la_k) w_k (\z) \epp
\end{equation}
Then
\begin{subequations}
\label{chimp}
\begin{align} \label{chim}
     \mspace{-9.mu}
     \chi (\la - 2 \i, \z) & = \frac{1}{\la - \z}
        - \frac{\Af (\z)}{\la - \z - 2 \i} 
        + \frac{\af (\z)}{\la - \z - 4 \i} 
        + \sum_{k=1}^N \biggl[ \frac{w_k (\z)}{\la - \la_k - 4 \i}
                               - \frac{w_k (\z)}{\la - \la_k} \biggr]
          \epc \\[2ex] \label{chip}
     \mspace{-9.mu}
     \chi (\la + 2 \i, \z) & = \frac{\fa (\z)}{\la - \z}
        - \frac{\Af (\z)}{\la - \z + 2 \i} 
        + \frac{1}{\la - \z + 4 \i} 
        + \sum_{k=1}^N \biggl[ \frac{w_k (\z)}{\la - \la_k}
                               - \frac{w_k (\z)}{\la - \la_k + 4 \i}
                              \biggr].
\end{align}
\end{subequations}
Let $\CR = \{ z \in {\mathbb C}| 1 < |\Im z| < 2 \}$. Then $\CR$ contains
all Bethe roots. The two functions $\chi (\la \mp 2 \i, \z_j)$ are
meromorphic in $\CR$. Their only poles inside $\CR$ are all simple and
are located at the Bethe roots and at $\z_j$. The corresponding residua are
\begin{subequations}
\begin{align}
     & \res_{\la = \la_k} \chi (\la - 2 \i, \z_j) = - w_k (\z_j) \epc \qd
       \res_{\la = \z_j} \chi (\la - 2 \i, \z_j) = 1 \epc \\[2ex]
     & \res_{\la = \la_k} \chi (\la + 2 \i, \z_j) = w_k (\z_j) \epc \qd
       \res_{\la = \z_j} \chi (\la + 2 \i, \z_j) = \fa (\z_j) \epp
\end{align}
\end{subequations}
Define two simple contours $\G^\pm$, such that (i) $\G^+$ is inside the
upper strip of $\CR$ and $\G^-$ is inside the lower strip of $\CR$, and
(ii) all Bethe roots and all $\z_j$, $j = 1, \dots, 2m$, are inside
$\G = \G^+ + \G^-$ (see figure~\ref{fig:spin1Gamma}).
Decompose $\G$ in such a way that $\G = {\cal B} +
{\cal I}$, where $\cal B$ contains only Bethe roots and $\cal I$ contains
only inhomogeneity parameters. Then we are very much in the same situation
as in \cite{GHS05}, and using the functions $\chi (\la \pm 2 \i, \z_j)$
we can transform the right hand side of (\ref{dens2x1}) into a multiple
integral over $\G$. As we shall see the notation
\begin{equation}
     g_j (\om|\z) = \begin{cases}
                       \chi (\om - 2 \i, \z) & j \le p \\
                       \chi (\om + 2 \i, \z) & j > p
                    \end{cases}              
\end{equation}
will prove useful in that exercise. Using also (\ref{zetaprod}) we obtain
\begin{multline} \label{dens2x2}
     D^{[2]} (\xv| \z)
        \biggl[ \prod_{j=1}^m - 2 \i \Bf(\x_j - \i) \sign \de_j \biggr]
        \prod_{1 \le j < k \le m} (\x_k - \x_j)^2 [(\x_k - \x_j)^2 + 4] \\
        = \lim_{\e \rightarrow 0+}
     \sum_{(\{\e^+\},\{\e^-\}) \in p_2 ({\mathbb Z}_{2m})} \:
     \biggl[ \prod_{j=1}^n \int_{\cal I} \frac{\rd \om_{\e_j^-}}{2 \p \i}
             \biggr]
     \sum_{\ell_{\e_1^+}, \dots, \ell_{\e_{2m-n}^+} = 1}^N \qd
     \sum_{\substack{(\{\de^+\}, \{\de^-\}) \in p_2 ({\mathbb Z}_{2m}) \\
                     \card \{\de^-\} = n}} \sign(PQ) \\[1ex]
     \underbrace{\sum_{R \in {\mathfrak S}^n} \sign (R) \:
                 g_{\e_1^-} (\om_{\e_1^-}, \z_{\de_{Rj}^-}) \dots
                 g_{\e_n^-} (\om_{\e_n^-}, \z_{\de_{Rn}^-})
                 }_{\det [ g_{\e_j^-} (\om_{\e_j^-}, \z_{\de_{Rk}^-})]}
     \frac{\det\bigl[w_{\ell_{\e_j^+}} (\z_{\de_k^+})\bigr]}
	  {\prod_{1 \le j < k \le 2m} (\om_j - \om_k - 2\i)}
           \biggr|_{\om_{\e_j^+} = \la_{\ell_{\e_j^+}}} \\[1ex]
     \Biggl[ \prod_{\substack{j = 1\\ \e_j^+ \le p}}^{2m-n}
	     - \F_{\e_j^+} (\la_{\ell_{\e_j^+}}) \Biggr]
     \Biggl[ \prod_{\substack{j = 1\\ \e_j^+ > p}}^{2m-n}
             \Fq_{\e_j^+} (\la_{\ell_{\e_j^+}}) \Biggr]
     \Biggl[ \prod_{\substack{j = 1\\ \e_j^- \le p}}^n
             \F_{\e_j^-} (\om_{\e_j^-}) \Biggr]
     \Biggl[ \prod_{\substack{j = 1\\ \e_j^- > p}}^n
             \Fq_{\e_j^-} (\om_{\e_j^-}) \Biggr] \displaybreak[0] \\[1ex]
        = \lim_{\e \rightarrow 0+}
     \underbrace{
     \sum_{(\{\e^+\},\{\e^-\}) \in p_2 ({\mathbb Z}_{2m})} \:
     \biggl[ \prod_{j=1}^{2m - n} \int_{\cal B}
             \frac{\rd \om_{\e_j^+}}{2 \p \i} \biggr]
     \biggl[ \prod_{j=1}^n \int_{\cal I} \frac{\rd \om_{\e_j^-}}{2 \p \i}
             \biggr]}_{= \prod_{j=1}^{2m} \int_\G \frac{\rd \om_j}{2 \p \i}}
     \frac{1}{\prod_{1 \le j < k \le 2m} (\om_j - \om_k - 2\i)} \\[1ex]
     \underbrace{
     \sum_{\substack{(\{\de^+\}, \{\de^-\}) \in p_2 ({\mathbb Z}_{2m}) \\
                     \card \{\de^-\} = n}} \sign(PQ)
     \det [ g_{\e_j^+} (\om_{\e_j^+}, \z_{\de_{Rk}^+})]
     \det [ g_{\e_j^-} (\om_{\e_j^-}, \z_{\de_{Rk}^-})]
     }_{= \det [g_j (\om_j, \z_k)]} \\[1ex]
     \Biggl[ \prod_{\substack{j = 1\\ \e_j^+ \le p}}^{2m-n}
	     \F_{\e_j^+} (\om_{\e_j^+}) \Biggr]
     \Biggl[ \prod_{\substack{j = 1\\ \e_j^+ > p}}^{2m-n}
             \Fq_{\e_j^+} (\om_{\e_j^+}) \Biggr]
     \Biggl[ \prod_{\substack{j = 1\\ \e_j^- \le p}}^n
             \F_{\e_j^-} (\om_{\e_j^-}) \Biggr]
     \Biggl[ \prod_{\substack{j = 1\\ \e_j^- > p}}^n
             \Fq_{\e_j^-} (\om_{\e_j^-}) \Biggr] \displaybreak[0] \\[1ex]
        = \lim_{\e \rightarrow 0+}
     \biggl[
     \prod_{j=1}^p \int_\G \frac{\rd \om_j}{2 \p \i} \F_j (\om_j) \biggr]
     \biggl[
     \prod_{j=p+1}^{2m} \int_\G \frac{\rd \om_j}{2 \p \i}
            \Fq_j (\om_j) \biggr]
     \frac{\det [g_j (\om_j, \z_k)]}
          {\prod_{1 \le j < k \le 2m} (\om_j - \om_k - 2\i)} \epp
\end{multline}
Note that we used the Laplace expansion formula for determinants in
the third equation.

To summarize up to this point, we have derived the equation
\begin{multline} \label{d2xnarrow}
     D^{[2]} (\xv| \z) =
        \lim_{\e \rightarrow 0+}
        \prod_{1 \le j < k \le m}
           \frac{1}{(\x_k - \x_j)^2 [(\x_k - \x_j)^2 + 4]} \\[1ex]
     \biggl[
     \prod_{j=1}^p \int_\G \frac{\rd \om_j}{2 \p \i} \F_j (\om_j) \biggr]
     \biggl[
     \prod_{j=p+1}^{2m} \int_\G \frac{\rd \om_j}{2 \p \i}
            \Fq_j (\om_j) \biggr]
     \frac{\det [g_j (\om_j, \z_k)]
           \prod_{j=1}^m \frac{\i \sign \de_j}{2 \Bf(\x_j - \i)}}
          {\prod_{1 \le j < k \le 2m} (\om_j - \om_k - 2\i)} \epp
\end{multline}
Here the factor $\prod_{j=1}^m \sign \de_j$ can be used to reorder the
columns in the determinant. Defining the alternating pattern
\begin{equation}
     \n_{2k - 1} = \x_k + \i (1 + \e) \epc \qd 
     \n_{2k} = \x_k - \i (1 + \e)
\end{equation}
and
\begin{equation}
     \chi^{(p, \e)}_{jk} = \begin{cases}
                          \chi(\la_j - 2 \i, \n_k) &
                          j = 1, \dots, p \\[1ex]
                          \chi(\la_j + 2 \i, \n_k) &
                          j = p + 1, \dots, 2m \epc
                       \end{cases}
\end{equation}
we find that
\begin{equation}
     \det [g_j (\om_j, \z_k)] \prod_{j=1}^m \sign \de_j
        = \det \chi^{(p, \e)}_{jk} \epp
\end{equation}

Note that the limit $\e \rightarrow 0+$ is not obvious at this stage,
because in the limit the poles of $g$ at the inhomogeneity parameters
unavoidably cross the narrow contour $\G$. Below we shall widen the
contour, while carefully taking account of the additional terms generated
during this process. As we shall see, all additional terms are of order
$\e$ and vanish in the limit.
\subsubsection{Fusion of wave functions}
Before coming to this point we have to recall that for the spin-1
density matrix elements we do not exactly need $ D^{[2]} (\xv| \z)$,
but certain combinations of these coefficients. This leeds to `fusion
of the wave functions' $\F$, $\Fq$, to be described in this subsection.

Let us consider a specific matrix element
\begin{equation} \label{d2symb}
     {D^{[2]}}^{\a_1, \dots, \a_m}_{\be_1, \dots, \be_m} (\x) =
        \frac{\<\Ps_0| {T^{[2]}}^{\a_1}_{\be_1} (\x_1) \dots
                       {T^{[2]}}^{\a_m}_{\be_m} (\x_m) |\Ps_0\>}
             {\<\Ps_0|\Ps_0\> \La^{[2]} (\x_1) \dots \La^{[2]} (\x_m)}
\end{equation}
of the spin-1 density matrix. Since the local space is spin-1, the
indices take three different values, $\a_j, \be_k = +, 0, -$. The right
hand side of (\ref{d2symb}) can be written as a linear combination of
coefficients $D^{[2]} (\xv|\x)$ which can be identified by means
of~(\ref{t1form}).

To begin with let us assume that ${T^{[2]}}^+_+ (\x_\ell) = {T^{[1]}}^+_+
(\x_\ell - \i) {T^{[1]}}^+_+ (\x_\ell + \i)$ is contained in the sequence
of mono\-dromy matrix elements on the right hand side of (\ref{d2symb}).
Then we must have $x_j = 2 \ell$ and $x_{j+1} = 2 \ell - 1$ for some $j \in
\{1, \dots, p - 1\}$ in all coefficients $D^{[2]} (\xv|\x)$ contained
in the linear combination for that specific density matrix element. Also
$\de_\ell = - 1 - \e$, and a factor
\begin{align} \label{aa}
     \prod_{k=1}^{2\ell - 1} & (\om_j - \z_k - 2\i)
        \prod_{k = 2\ell + 1}^{2m} (\om_j - \z_k) 
     \prod_{k=1}^{2\ell - 2} (\om_{j+1} - \z_k - 2\i)
        \prod_{k = 2\ell}^{2m} (\om_{j+1} - \z_k) \notag \\[1ex]
     & = \bigl[ (\om_j - \x_\ell - \i)(\om_{j+1} - \x_\ell - \i)
	  - \i \e (\om_j - \om_{j+1} - 2 \i) + \e(2 + \e) \bigr] \notag \\
     & \qqd \times \prod_{n=j}^{j+1} \: \prod_{k=1}^{\ell - 1}
         \bigl[(\om_n - \x_k - 3\i) (\om_n - \x_k - \i)
	       + \e(2 + \e) \bigr] \notag \\ & \mspace{180.mu}
       \times \prod_{k = \ell + 1}^m
         \bigl[(\om_n - \x_k - \i) (\om_n - \x_k + \i)
	       + \e(2 + \e) \bigr] \notag \\[1ex]
     & = F_\ell (\om_j) F_\ell (\om_{j+1}) + {\cal O} (\e)
\end{align}
appears. Here we used the `spin-1 wave function' $F_\ell$ defined in
(\ref{d2wave}).

In a similar way we may consider all the matrix elements of $T^{[2]}$
using for simplification the right hand side of (\ref{t1form}). E.g.\
if ${T^{[2]}}^+_0 (\x_\ell) = \sqrt{2} {T^{[1]}}^+_+ (\x_\ell + \i)
{T^{[1]}}^+_- (\x_\ell - \i)$ is contained in the sequence of mono\-dromy
matrix elements on the right hand side of (\ref{d2symb}), then $\de_\ell
= 1 + \e$, and we have $x_j = 2 \ell$, $x_{j+1} = 2\ell - 1$ for some
$j \in \{1, \dots, p - 1\}$ and $x_i = 2 \ell$ for some $i \in
\{p + 1, \dots, 2m\}$. Thus, there is a factor
\begin{align} \label{ab}
     & \sqrt{2} \bigl[ (\om_j - \x_\ell - \i)(\om_{j+1} - \x_\ell - \i)
         + \i (2 + \e) (\om_j - \om_{j+1} - 2 \i) + \e(2 + \e) \bigr]
	 \notag \\
     & \qqd \times \prod_{n=j}^{j+1} \: \prod_{k=1}^{\ell - 1}
         \bigl[(\om_n - \x_k - 3\i) (\om_n - \x_k - \i)
	       + \e(2 + \e) \bigr] \notag \\ & \mspace{180.mu}
       \times \prod_{k = \ell + 1}^m
         \bigl[(\om_n - \x_k - \i) (\om_n - \x_k + \i)
	       + \e(2 + \e) \bigr] \notag \displaybreak[0] \\[1ex]
     & \qqd \times (\om_i - \x_\ell + \i - \i \e) \prod_{k=1}^{\ell - 1}
         \bigl[(\om_i - \x_k + 3\i) (\om_i - \x_k + \i)
	       + \e(2 + \e) \bigr] \notag \\ & \mspace{180.mu}
       \times \prod_{k = \ell + 1}^m
         \bigl[(\om_i - \x_k - \i) (\om_i - \x_k + \i)
	       + \e(2 + \e) \bigr] \notag \\[1ex]
     & \equiv \sqrt{2}\, F_\ell (\om_j) F_\ell (\om_{j+1})
              \overline{F}_\ell (\om_i) + {\cal O} (\e)
\end{align}
under the integral. Again $F_\ell$ and $\overline{F}_\ell$ are taken from
(\ref{d2wave}). We use the notation `$\equiv$' for `equal under
the multiple integral' (\ref{d2xnarrow}). The crucial point here is that
the term proportional to $(2 + \e) (\om_j - \om_{j+1} - 2 \i)$ does not
contribute under the integral (\ref{d2xnarrow}) for symmetry
considerations. This is because it multiplies a function which is
symmetric in $\om_j$ and $\om_{j+1}$ in (\ref{ab}) and the only other
terms under the integral depending on $\om_j$ and $\om_{j+1}$ are
$\det [g_j (\om_j, \z_k)]$ and $\prod_{1 \le j < k \le 2m}
(\om_j - \om_k - 2\i)$. But
\[
     \frac{(\om_j - \om_{j+1} - 2 \i) \det [g_j (\om_j, \z_k)]}
          {\prod_{1 \le j < k \le 2m} (\om_j - \om_k - 2\i)}
\]
is antisymmetric in $\om_j$ and $\om_{j+1}$.

Another example is the matrix element ${T^{[2]}}^0_0 (\x_\ell) =
{T^{[1]}}^-_+ (\x_\ell - \i) {T^{[1]}}^+_- (\x_\ell + \i) +
{T^{[1]}}^-_- (\x_\ell - \i) {T^{[1]}}^+_+ (\x_\ell + \i)$ for which
$\de_\ell = - 1 - \e$. It is the only matrix element which we have to
express by a sum of two products of monodromy matrix elements with
spin-$\2$ auxiliary space, and it contributes a factor
\begin{align} \label{cbpda}
     \prod_{k=1}^{2\ell - 1} & (\om_j - \z_k - 2\i)
        \prod_{k = 2\ell + 1}^{2m} (\om_j - \z_k) \notag \\ &
     \times \biggl[
     \prod_{k=1}^{2\ell - 1} (\om_i - \z_k + 2\i)
        \prod_{k = 2\ell + 1}^{2m} (\om_i - \z_k) +
     \prod_{k=1}^{2\ell - 2} (\om_i - \z_k + 2\i)
        \prod_{k = 2\ell}^{2m} (\om_i - \z_k) \biggr] \notag \\[1ex] &
     = 2 F_\ell (\om_j) \overline{F}_\ell (\om_i) + {\cal O} (\e) \epc
\end{align}
where $j \in \{1, \dots, p\}$ and $i \in \{p + 1, \dots, 2m\}$.

\renewcommand{\arraystretch}{1.5}
\begin{table}[t]
\begin{minipage}{\linewidth}
    \centering
    \begin{tabular}{cl}
      \toprule
        Matrix element & Factor under the integral\\
      \midrule
        $ {T^{[2]}}^+_+ (\x_\ell)$ &
        $F_\ell (\om_j) F_\ell (\om_{j+1}) + {\cal O} (\e)$ \\
        $ {T^{[2]}}^+_0 (\x_\ell)$ &
        $\sqrt{2}\, F_\ell (\om_j) F_\ell (\om_{j+1})
         \overline{F}_\ell (\om_i) + {\cal O} (\e)$ \\
        $ {T^{[2]}}^+_- (\x_\ell)$ &
        $F_\ell (\om_j) F_\ell (\om_{j+1}) \overline{F}_\ell (\om_i)
         \overline{F}_\ell (\om_{i+1}) + {\cal O} (\e)$ \\
        $ {T^{[2]}}^0_+ (\x_\ell)$ &
        $\sqrt{2}\, F_\ell (\om_j) + {\cal O} (\e)$ \\
        $ {T^{[2]}}^0_0 (\x_\ell)$ &
        $2 F_\ell (\om_j) \overline{F}_\ell (\om_i) + {\cal O} (\e)$ \\
        $ {T^{[2]}}^0_- (\x_\ell)$ &
        $ \sqrt{2}\, F_\ell (\om_j) \overline{F}_\ell (\om_i)
         \overline{F}_\ell (\om_{i+1}) + {\cal O} (\e)$ \\
        $ {T^{[2]}}^-_+ (\x_\ell)$ & $1$ \\
        $ {T^{[2]}}^-_0 (\x_\ell)$ &
        $\sqrt{2}\; \overline{F}_\ell (\om_i) + {\cal O} (\e)$ \\
        $ {T^{[2]}}^-_- (\x_\ell)$ &
        $\overline{F}_\ell (\om_i)
         \overline{F}_\ell (\om_{i+1}) + {\cal O} (\e)$ \\
      \bottomrule
    \end{tabular}
    \caption{\label{tab:fusedf} The polynomials under the integral.}
\end{minipage}
\end{table}
\renewcommand{\arraystretch}{1.1}

Proceeding case by case in a similar way we obtain the result exposed
in tabular~\ref{tab:fusedf}. In the tabular it is always implied that
$j, j + 1 \in \{1, \dots, p\}$ and $i, i + 1 \in \{p+1, \dots, 2m\}$.
Inspecting the tabular we see that, up to corrections of the order of
$\e$, the `wave function' under the integral is composed in the
following way: For every zero in the sequences $(\a_j)$ and $(\be_k)$
we obtain a factor of $\sqrt{2}$, amounting to a total factor of
$2^{(n_0 (\a) + n_0 (\be)) /2} = 2^{m - n_+ (\a) - n_- (\be)}$. For
every plus in $(\a_j)$ a factor $F_\ell (\om_j) F_\ell (\om_{j+1})$
appears and for every zero a factor $F_\ell (\om_j)$. From the sequence
$(\be_k)$ we obtain a factor $\overline{F}_\ell (\om_i)$ for every zero
and a factor $\overline{F}_\ell (\om_i) \overline{F}_\ell (\om_{i+1})$
for every minus sign. This makes a total number of $2 n_+ (\a) + n_0 (\a)
+ n_0 (\be) + 2 n_- (\be) = 2m$ factors, and $p = 2 n_+ (\a) + n_0 (\a)$.
With the factors we obtain a sequence $(z_j)_{j=1}^{2m}$,
$z_j \in \{1, \dots, m\}$ by arranging them in the order of ascending
$\om_j$: $F_{z_1} (\om_1) \dots F_{z_p} (\om_p) \overline{F}_{z_{p+1}}
\dots \overline{F}_{z_{2m}} (\om_{2m})$.

Thus, we have obtained the representation
\begin{multline} \label{d2narrow}
     {D^{[2]}}^{\a_1, \dots, \a_m}_{\be_1, \dots, \be_m} (\x) =
           \frac{2^{- n_+ (\a) - n_- (\be)}
                 \prod_{j=1}^m \frac{\i}{\Bf(\x_j - \i)}}
                {\prod_{1 \le j < k \le m} (\x_k - \x_j)^2
                 [(\x_k - \x_j)^2 + 4]} \\[1ex] \lim_{\e \rightarrow 0+}
     \biggl[ \prod_{j=1}^{2m} \int_\G \frac{\rd \om_j}{2 \p \i} \biggr]
     \frac{\det \chi^{(p, \e)}_{jk}}
          {\prod_{1 \le j < k \le 2m} (\om_j - \om_k - 2\i)} \\[1ex]
     \bigl[F_{z_1} (\om_1) \dots F_{z_p} (\om_p)
           \overline{F}_{z_{p+1}} (\om_{p+1}) \dots
           \overline{F}_{z_{2m}} (\om_{2m}) + {\cal O} (\e) \bigr]
\end{multline}
for the spin-1 density matrix elements. What remains to do is to
calculate the limit $\e \rightarrow 0$. For this purpose we have
to deform the integration contours first.

\subsubsection{Widening the contours}
Next we would like to show that
\begin{multline} \label{d2narrowwide}
     \biggl[ \prod_{j=1}^{2m} \int_\G \frac{\rd \om_j}{2 \p \i} \biggr]
     \frac{\det \chi^{(p, \e)}_{jk}}
          {\prod_{1 \le j < k \le 2m} (\om_j - \om_k - 2\i)} \\[.5ex]
     \bigl[F_{z_1} (\om_1) \dots F_{z_p} (\om_p)
           \overline{F}_{z_{p+1}} (\om_{p+1}) \dots
           \overline{F}_{z_{2m}} (\om_{2m}) + {\cal O} (\e) \bigr] \\[1ex]
     \mspace{-120.mu} =
     \biggl[ \prod_{j=1}^p \int_{\cal C}
                              \frac{\rd \om_j}{2 \p \i} \biggr]
     \biggl[ \prod_{j = p + 1}^{2m} \int_{\overline{\cal C}}
                              \frac{\rd \om_j}{2 \p \i} \biggr]
     \frac{\det \chi^{(p, \e)}_{jk}}
          {\prod_{1 \le j < k \le 2m} (\om_j - \om_k - 2\i)} \\[.5ex]
     \bigl[F_{z_1} (\om_1) \dots F_{z_p} (\om_p)
           \overline{F}_{z_{p+1}} (\om_{p+1}) \dots
           \overline{F}_{z_{2m}} (\om_{2m}) + {\cal O} (\e) \bigr]
           + {\cal O} (\e) \epp
\end{multline}
We recall that the simple closed contours $\cal C$, $\overline{\cal C}$
are defined in such a way that
\begin{align} \label{contrels}
     & \CC = \CC^+ + \CC^- \epc \qd \CCq = \CCq^+ + \CCq^- \notag \\
     & \CC^\pm \epc \CCq^\pm \subset \CS^\pm \notag \\
     & \CC^- = \CCq^- \epc \qd \CC^-\ \text{inside}\ \CC^+ - 2\i \epc
       \qd \CCq^+ - 2\i\ \text{inside}\ \CC^- \notag \\
     & \G\ \text{inside}\ \CC \epc \CCq \epp
\end{align}
For $\CC^\pm$, $\CCq^\pm$ we may take large rectangles inside $\CS^\pm$
which are slightly narrower than 2 in imaginary direction. The third
line in (\ref{contrels}) is a closed contour analogon of the
regularization by infinitesimal shifts of the contours in
\cite{Kitanine01}. The Bethe roots of the dominant state come in complex
conjugated pairs, so-called 2-strings. For their enumeration we shall
employ the same convention as in \cite{Kitanine01}. Those in the upper
half plane will be labeled by odd integers and those in the lower half
plane by even integers. By definition the contours $\CC$ and $\CCq$
encircle not only all Bethe roots $\la_j$ and all inhomogeneities $\n_j$
but also the down-shifted upper Bethe roots $\la_{2j-1} - 2\i$ and the
up-shifted lower Bethe roots $\la_{2j} + 2\i$ as well as the down-shifted
upper inhomogeneities $\n_{2j-1} - 2\i$ and the up-shifted lower
inhomogeneities $\n_{2j} + 2\i$.

In preparation of the proof of (\ref{d2narrowwide}) we introduce the
notation
\begin{equation}
      f(\om_1, \dots, \om_{2m}) =
         \frac{F_{z_1} (\om_1) \dots F_{z_p} (\om_p)
               \overline{F}_{z_{p+1}} (\om_{p+1}) \dots
               \overline{F}_{z_{2m}} (\om_{2m})+ {\cal O} (\e)}
              {\prod_{1 \le j < k \le 2m} (\om_j - \om_k - 2\i)} \epc
\end{equation}
where the polynomial, including the ${\cal O} (\e)$ contribution, is the
same as under the integrals in (\ref{d2narrowwide}). Then the integral
on the right hand side of (\ref{d2narrowwide}) can be written as
\begin{multline} \label{startp}
     \sum_{Q \in {\mathfrak S}^{2m}} \sign (Q)
        \int_{\CC} \frac{\rd \om_1}{2 \p \i}
        \chi(\om_1 - 2\i, \n_{Q1}) \dots
        \int_{\CC} \frac{\rd \om_p}{2 \p \i}
        \chi(\om_p - 2\i, \n_{Qp}) \\
        \int_{\CCq} \frac{\rd \om_{2m}}{2 \p \i}
        \chi(\om_{2m} + 2\i, \n_{Q(2m)}) \dots
        \int_{\CCq} \frac{\rd \om_{p+1}}{2 \p \i}
        \chi(\om_{p+1} + 2\i, \n_{Q(p+1)}) \:
        f(\om_1, \dots, \om_{2m}) \epp
\end{multline}
We shall show that, if we successively replace the integrals in this
expression by integrals over $\G$, the total error will be of the order
$\e$.

(a) For the rightmost integral we note that $f$ considered as a function
of $\om_{p+1}$ is holomorphic inside $\CCq$. There is a factor $(\om_1 -
\om_{p+1} - 2\i) \dots (\om_p - \om_{p+1} - 2\i) (\om_{p+1} - \om_{p+2}
- 2\i) \dots (\om_{p+1} - \om_{2m} - 2\i)$ in the denominator, but with
our choice of contours $\om_j - 2\i$ is outside $\CCq$ for $\om_j \in
\CC$, $j = 1, \dots, p$, and the same is true for $\om_j + 2\i$ for
$\om_j \in \CCq$, $j = p + 2, \dots, 2m$. The function $\chi (\om_{p+1}
+ 2\i, \n_{Q(p+1)})$ has outside $\G$ but inside $\CCq$ at most a single
pole occurring at $\n_{Q(p+1)} - 2\i$ if $Q(p+1)$ is odd. Then there is
$\ell \in {\mathbb Z}_m$ such that $\n_{Q(p+1)} - 2\i = \x_\ell - \i +
\i \e$ and, if we contract the contour from $\CCq$ to $\G$, the pole
contributes a term having a factor $\overline{F}_{z_{p+1}}
(\x_\ell - \i + \i \e) = {\cal O} (\e)$ in the numerator. Hence, the
numerator of this term is ${\cal O} (\e)$. In the denominator we have
factors of $\om_j - \n_{Q(p+1)} = \om_j - \x_\ell - \i - \i \e$ for
$j = 1, \dots, p$ or $\n_{Q(p+1)} - \om_j - 4 \i = \x_\ell - \om_j - 3 \i
+ \i \e$ for $j = p + 2, \dots, 2m$. It follows that the absolute
value of the denominator is bounded from below if the $\om_j$ are on
$\CC$ for $j = 1, \dots, p$ or on $\CCq$ for $j = p+2, \dots, 2m$. Hence,
the additional term that may be generated by contracting the contour from
$\CCq$ to $\G$ will at most contribute to order $\e$, even after
performing the summation and the remaining integrations in (\ref{startp}).
We symbolize this by writing
\begin{multline}
     \int_{\CCq} \frac{\rd \om_{p+1}}{2 \p \i}
        \chi(\om_{p+1} + 2\i, \n_{Q(p+1)}) f(\om_1, \dots, \om_{2m})
        \equiv_\e \\
     \int_\G \frac{\rd \om_{p+1}}{2 \p \i}
        \chi(\om_{p+1} + 2\i, \n_{Q(p+1)}) f(\om_1, \dots, \om_{2m}) \epp
\end{multline}

(b) In order to proceed by induction we define
\begin{subequations}
\begin{align}
     & I_0 = f(\om_1, \dots, \om_{2m}) \epc \\[1ex]
     & I_n = \int_\G \frac{\rd \om_{p+n}}{2 \p \i}
        \chi(\om_{p+n} + 2\i, \n_{Q(p+n)}) I_{n-1} \epc \qd
        n = 1, \dots, 2m - p \epp
\end{align}
\end{subequations}
We want to show that
\begin{equation} \label{indhypbar}
     \int_{\CCq} \frac{\rd \om_{p+n}}{2 \p \i}
        \chi(\om_{p+n} + 2\i, \n_{Q(p+n)}) I_{n-1} \equiv_\e I_n \epp
\end{equation}
We have already shown that this is valid for $n=1$. To proceed further
we have to know the structure of $I_n$.

(c) As a preparatory step let us consider the left hand side
(\ref{indhypbar}) for $n=2$. Then
\begin{multline} \label{i1}
     I_1 = \sum_{k_{p+1} = 1}^{N} f(\om_1, \dots, \om_p, \la_{k_{p+1}},
              \om_{p+2}, \dots, \om_{2m}) w_{k_{p+1}} (\n_{Q(p+1)}) \\ +
              f(\om_1, \dots, \om_p, \n_{Q(p+1)}, \om_{p+2}, \dots,
                \om_{2m}) \fa (\n_{Q(p+1)}) \epp
\end{multline}
Here we have used that $f$ is holomorphic as a function of $\om_{p+1}$
for $\om_{p+1}$ on and inside $\G$. We insert $I_1$ into the left hand
side of (\ref{indhypbar}) for $n=2$ and contract the integration contour
from $\CCq$ to $\G$. Due to our special choice of the contours $\CC$,
$\CCq$ the poles at $\om_{p+2} = \om_j - 2\i$, $j = 1, \dots, p$ and
at $\om_{p+2} = \om_j + 2\i$, $j = p + 3, \dots, 2m$, remain outside
the contour during the process of deformation. The only pole of
$\chi(\om_{p+2} + 2\i, \n_{Q(p+2)})$ which may be crossed is, in case
that $Q(p+2)$ is odd, a simple pole at $\n_{Q(p+2)} - 2\i$. The situation
is the same as in case (a) above. And as above we can see that such a
term gives only an order-$\e$ contribution, even after performing the sum
and the remaining integrals: $\n_{Q(p+2)} = \x_\ell + \i + \i \e\ \then\
\overline{F}_{z_{p+2}} (\n_{Q(p+2)} - 2\i) = {\cal O} (\e)$ and in the
denominator we may have $\om_j - \n_{Q(p+2)} = \om_j - \x_\ell - \i -
\i \e$ for $j = 1, \dots, p$ or $\n_{Q(p+2)} - \om_j - 4\i =
\x_\ell - \om_j - 3\i + \i \e$ for $j = p + 3, \dots, 2m$, as before, or
$\la_{k_{p+1}} - \n_{Q(p+2)}$ or $\n_{Q(p+1)} - \n_{Q(p+2)}$. The latter
terms are of no danger, since we assume that the $\x_j$ are mutually
distinct and distinct from all Bethe roots. Thus, we see that the same
argument as above works.

However, we now have singularities of $I_1$ which are crossed in the
course of the deformation of the contour. They give additional
contributions. The summand with $k_{p+1} = j$ has a term $\la_j -
\om_{p+2} - 2\i$ in the denominator, giving rise to a simple pole in
$\om_{p+2}$ at $\la_j - 2\i$, if $j$ is odd. When contracting the
contour of the $\om_{p+2}$-integral this pole causes a contribution
(see (\ref{defchi}), (\ref{defw})) proportional to
\begin{equation}
     \chi(\la_j,\n_{Q(p+2)}) w_j (\n_{Q(p+1)}) =
        \fa' (\la_j) w_j (\n_{Q(p+2)}) w_j (\n_{Q(p+1)})
\end{equation}
which is symmetric in $p+1$, $p+2$. Such a term vanishes under the sum
over all permutations in (\ref{startp}). Another additional contribution
comes from the second term in (\ref{i1}), which has a factor $\n_{Q(p+1)}
- \om_{p+2} - 2\i$ in the denominator. So we have a pole at $\n_{Q(p+1)}
- 2\i$. It will give only ${\cal O} (\e)$-corrections when we integrate
over $\om_{p+2}$, as we have already seen.

(d) In the general case the argument is very similar. $I_{n-1}$ is
obtained by iterating the integrations over $\G$. In every integration
the $I_{j-1}$ under the integral is holomorphic on and inside $\G$ by
construction. Hence, we obtain a sum over the pole contributions of
$\chi(\om_{p+j} + 2\i, \n_{Q(p+j)})$. This means that $I_{n-1}$ is
a sum over terms of the form
\begin{equation} \label{term}
     t(\om_{p+n}) = f(\om_1, \dots, \om_p, x_{p+1}, \dots, x_{p+n-1},
                      \om_{p+n}, \dots, \om_{2m}) \:
                      y_{p+1} \dots y_{p+n-1} \epc
\end{equation}
where $x_j$ is either a Bethe root or $\n_{Qj}$. If $x_j$ is a Bethe root,
say $\la_\ell$, then the corresponding factor $y_j = w_\ell (\n_{Qj})$,
if $x_j = \n_{Qj}$, then $y_j = \fa (\n_{Qj})$. Let us consider
\begin{equation}
     \int_{\CCq} \frac{\rd \om_{p+n}}{2 \p \i}
        \chi(\om_{p+n} + 2\i, \n_{Q(p+n)}) t(\om_{p+n}) \epp
\end{equation}
If we shrink the contour to $\G$, we obtain at most one pole contribution
from $\chi$ which is at $\n_{Q(p+n)} - 2\i$ if $Q(p+n)$ is odd. This
contribution is ${\cal O} (\e)$ by the same argument as above. Also
for the contributions stemming from $t$ we can argue as above. Note that
we do not have to consider double poles (or poles of even higher order),
since, if $x_j$ and $x_k$ are the same Bethe root, $\la_\ell$ say, then
$t(\om_{p+n})$ has a factor $w_\ell (\n_{Qj}) w_\ell (\n_{Qk})$, which
vanishes under the antisymmetrizing sum in (\ref{startp}). Thus, we
have accomplished the proof of (\ref{indhypbar}).

(e) It follows from (\ref{indhypbar}) that we can replace all the
$\CCq$-integrals in (\ref{startp}) by $\G$-integrals. The total
error will be only of order $\e$. To finish the proof of
(\ref{d2narrowwide}) we have to proceed with the contours $\CC$.
Let us set
\begin{subequations}
\begin{align}
     & J_{p+1} = I_{2m - p} \epc \\[1ex]
     & J_n = \int_\G \frac{\rd \om_n}{2 \p \i}
        \chi(\om_n - 2\i, \n_{Qn}) J_{n+1} \epc \qd
        n = p, \dots, 1 \epp
\end{align}
\end{subequations}
We want to show that
\begin{equation} \label{indhyp}
     \int_{\CC} \frac{\rd \om_n}{2 \p \i}
        \chi(\om_n - 2\i, \n_{Qn}) J_{n+1} \equiv_\e J_n \epp
\end{equation}
The proof is very similar as before.

(f) Let us start with $n=p$. Then $J_{p+1} = I_{2m-p}$ is a sum over
terms of the form
\begin{equation} \label{serm}
     s(\om_p) = f(\om_1, \dots, \om_p, x_{p+1}, \dots, x_{2m})
                y_{p+1} \dots y_{2m} \epc
\end{equation}
as follows from (\ref{term}). If we shrink the contour in
\begin{equation}
     \int_{\CC} \frac{\rd \om_p}{2 \p \i}
        \chi(\om_p - 2\i, \n_{Qp}) s(\om_p)
\end{equation}
to $\G$, we obtain at most a single pole contribution from $\chi$
stemming from a pole at $\n_{Qp} + 2\i$ in case that $Qp$ is even. Then
$\n_{Qp} + 2\i = \x_\ell + \i - \i \e$ for some $\ell \in {\mathbb Z}_m$.
But now $F_{z_p} (\x_\ell + \i - \i \e) = {\cal O} (\e)$, and the numerator
in the generated term is ${\cal O} (\e)$. The denominator contains
terms $\om_j - (\n_{Qp} + 2\i) - 2\i = \om_j - \x_\ell - 3\i + \i \e$
which are bounded from below in the absolute value for $\om_j \in \CC$,
$j = 1, \dots, p - 1$. Hence, the whole term is of order $\e$, even
after integration and summation, and can safely be forgotten. As for the
singularities of $s(\om_p)$ we have again two types. If $x_j$ is a
Bethe root, say $\la_\ell$, then a term $\om_p - \la_\ell - 2\i$ occurs
in the denominator. It comes together with a factor $w_\ell (\n_{Qj})$.
When calculating the residue at $\la_\ell + 2\i$, which is non-zero
only if $\ell$ is even, we obtain something proportional to
\begin{equation}
     \chi(\la_\ell,\n_{Qp}) w_\ell (\n_{Qj}) =
        \fa' (\la_\ell) w_\ell (\n_{Qp}) w_\ell (\n_{Qj})
\end{equation}
which yields a term that vanishes under the sum in (\ref{startp}) due
to symmetry reasons. Double poles can be excluded by the same argument
as above. Finally we may have $x_j = \n_{Qj}$. Then a factor $\om_p
- \n_{Qj} - 2\i$ is present in the denominator resulting again at most
in an ${\cal O} (\e)$-contribution.

(g) Iterating the above arguments we obtain (\ref{indhyp}), and the proof
of (\ref{d2narrowwide}) is complete. With (\ref{d2narrowwide}) we can
now perform the limit $\e \rightarrow 0+$, because it is trivial at
the right hand side of the equation. With the definition
\begin{equation}
     \chi^{(p)}_{jk} = \lim_{\e \rightarrow 0+} \chi^{(p, \e)}_{jk}
\end{equation}
we obtain
\begin{multline} \label{epszero}
     \lim_{\e \rightarrow 0+}
     \biggl[ \prod_{j=1}^{2m} \int_\G \frac{\rd \om_j}{2 \p \i} \biggr]
     \frac{\det \chi^{(p, \e)}_{jk}}
          {\prod_{1 \le j < k \le 2m} (\om_j - \om_k - 2\i)} \\[.5ex]
     \bigl[F_{z_1} (\om_1) \dots F_{z_p} (\om_p)
           \overline{F}_{z_{p+1}} (\om_{p+1}) \dots
           \overline{F}_{z_{2m}} (\om_{2m}) + {\cal O} (\e) \bigr] \\[1ex]
     \mspace{-120.mu} =
     \biggl[ \prod_{j=1}^p
             \int_{\cal C} \frac{\rd \om_j}{2 \p \i}
             F_{z_j} (\om_j) \biggr]
     \biggl[ \prod_{j = p + 1}^{2m} 
             \int_{\overline{\cal C}} \frac{\rd \om_j}{2 \p \i}
             \overline{F}_{z_j} (\om_j) \biggr]
     \frac{\det \chi^{(p)}_{jk}}
          {\prod_{1 \le j < k \le 2m} (\om_j - \om_k - 2\i)} \epp
\end{multline}

With this we have derived the multiple integral representation
\begin{multline} \label{d2wide}
     {D^{[2]}}^{\a_1, \dots, \a_m}_{\be_1, \dots, \be_m} (\x) =
           \frac{2^{- n_+ (\a) - n_- (\be)}}
                {\prod_{1 \le j < k \le m} (\x_k - \x_j)^2
                 [(\x_k - \x_j)^2 + 4]} \\[1ex]
     \biggl[ \prod_{j=1}^p
             \int_{\cal C} \frac{\rd \om_j}{2 \p \i}
             F_{z_j} (\om_j) \biggr]
     \biggl[ \prod_{j = p + 1}^{2m} 
             \int_{\overline{\cal C}} \frac{\rd \om_j}{2 \p \i}
             \overline{F}_{z_j} (\om_j) \biggr]
     \frac{\det \chi^{(p)}_{jk} \: \prod_{j=1}^m \frac{\i}{\Bf(\x_j - \i)}}
          {\prod_{1 \le j < k \le 2m} (\om_j - \om_k - 2\i)}
\end{multline}
for the inhomogeneous density matrix of the isotropic spin-1 chain.

\subsection{The linear integral equations}
In this subsection we shall derive a pair of coupled integral equations
for the functions $\chi(\la \pm 2\i, \n)$. For this purpose we first of
all note that the function $\chi$ has been defined in (\ref{defchi}) in
such a way that
\begin{equation} \label{chilaj}
     \chi (\la_j, \n) = \fa' (\la_j) w_j (\n) \epc \qd j = 1, \dots, N
\end{equation}
(see (\ref{defw})). We shall use some of the functions that appear
in the formulation of the thermodynamics of the model \cite{Suzuki99}
and that are collected in appendix~\ref{app:auxfun}. In first place
we need the functions $\Af$ and $\Afq$ (see (\ref{defas})).
It follows from their definition and from the fact that $\La^{[1]}$
has no zeros in $\CS^+ \cup \CS^-$ that the only zeros of $\Af (\la + 2\i)$
in $\CS^-$ are simple zeros at $\la_{2j - 1} - 2\i$, while the only
poles in $\CS^-$ are simple poles at $\la_{2j}$. Similarly, the only
zeros of $\Afq (\la - 2\i)$ in $\CS^+$ are simple zeros at
$\la_{2j} + 2\i$ and its only poles in $\CS^+$ are simple and located at
$\la_{2j - 1}$.

Hence, using (\ref{chip}), (\ref{chilaj}), we conclude that the only
singularities of the function $\chi(\la + 2\i, \x^+)/\Af (\la + 2 \i)$
inside $\CS^-$ are
\begin{enumerate}
\item
a simple pole at $\la = \x^-$ with residue $- 1$,
\item
simple poles at $\la = \la_{2k-1} - 2\i$ with residua $w_{2k-1} (\x^+)$.
\end{enumerate}
Similarly, all singularities of $\chi(\la - 2\i, \x^+)/\Afq (\la - 2\i)$
inside $\CS^+$ are (see (\ref{chim}) and (\ref{chilaj}))
\begin{enumerate}
\item
a simple pole at $\la = \x^+$ with residue $1/\Afq (\x^-)$,
\item
simple poles at $\la = \la_{2k} + 2\i$ with residua $- w_{2k} (\x^+)$.
\end{enumerate}
With this it follows by means of (\ref{chimp}) that
\begin{subequations}
\begin{align}
     & \chi(\la - 2\i, \x^+) = \frac{\Af (\x^+)}{\la - \x^+ - 4\i}
        - \frac{\Bf (\x^-)}{\Af (\x^-)} \frac{1}{\la - \x^+ - 2\i}
        - \frac{\af (\x^-)}{\Af (\x^-)} \frac{1}{\la - \x^-}
        \notag \\[1ex] & \mspace{18.mu}
        - \int_{\CC^+} \frac{\rd \m}{2 \p \i}
             \frac{\chi (\m - 2\i, \x^+)}{\Afq (\m - 2\i)}
             K(\la - \m)
        + \int_{\CCq^-} \frac{\rd \m}{2 \p \i}
             \frac{\chi (\m + 2\i, \x^+)}{\Af (\m + 2\i)} K(\la - \m - 4\i)
          \epc \\[2ex]
     & \chi(\la + 2\i, \x^+) = \frac{\Af (\x^+)}{\la - \x^+}
        - \frac{\Bf (\x^-)}{\Af (\x^-)} \frac{1}{\la - \x^-}
        - \frac{\af (\x^-)}{\Af (\x^-)} \frac{1}{\la - \x^- + 4\i}
        \notag \\[1ex] & \mspace{18.mu}
        - \int_{\CCq^+} \frac{\rd \m}{2 \p \i}
             \frac{\chi (\m - 2\i, \x^+)}{\Afq (\m - 2\i)}
             K(\la - \m + 4\i)
        + \int_{\CC^-} \frac{\rd \m}{2 \p \i}
             \frac{\chi (\m + 2\i, \x^+)}{\Af (\m + 2\i)} K(\la - \m)
\end{align}
\end{subequations}
for $\la \in \CC^+$ in the first equation and $\la \in \CC^-$ in the
second equation.

If now $\n = \x^-$ we have to repeat almost the same considerations.
The function $\chi(\la + 2\i, \x^-)/\Af (\la + 2 \i)$ has the following
singularities inside $\CS^-$:
\begin{enumerate}
\item
an simple pole at $\la = \x^-$ with residue $\af (\x^-)/\Af (\x^+)$,
\item
simple poles at $\la = \la_{2k-1} - 2\i$ with residues $w_{2k-1} (\x^-)$.
\end{enumerate}
Furthermore, all singularities of $\chi(\la - 2\i, \x^-)/\Afq (\la - 2\i)$
inside $\CS^+$ are
\begin{enumerate}
\item
a simple pole at $\x^+$ with residue $- \fa(\x^-)$,
\item
simple poles at $\la = \la_{2k} + 2\i$ with residua $- w_{2k} (\x^-)$.
\end{enumerate}
Thus, we obtain
\begin{subequations}
\begin{align}
     & \chi(\la - 2\i, \x^-) =
        - \frac{\af (\x^-)}{\Af (\x^+)} \frac{1}{\la - \x^+ - 4\i}
        - \frac{\Bf (\x^-)}{\Af (\x^+)} \frac{1}{\la - \x^+}
        + \frac{\Af (\x^-)}{\la - \x^-}
        \notag \\[1ex] & \mspace{18.mu}
        - \int_{\CC^+} \frac{\rd \m}{2 \p \i}
             \frac{\chi (\m - 2\i, \x^-)}{\Afq (\m - 2\i)} K(\la - \m)
        + \int_{\CCq^-} \frac{\rd \m}{2 \p \i}
             \frac{\chi (\m + 2\i, \x^-)}{\Af (\m + 2\i)} K(\la - \m - 4\i)
          \epc \\[2ex]
     & \chi(\la + 2\i, \x^-) =
        - \frac{\af (\x^-)}{\Af (\x^+)} \frac{1}{\la - \x^+}
        - \frac{\Bf (\x^-)}{\Af (\x^+)} \frac{1}{\la - \x^- + 2\i}
        + \frac{\Af (\x^-)}{\la - \x^- + 4\i}
        \notag \\[1ex] & \mspace{18.mu}
        - \int_{\CCq^+} \frac{\rd \m}{2 \p \i}
             \frac{\chi (\m - 2\i, \x^-)}{\Afq (\m - 2\i)}
             K(\la - \m + 4\i)
        + \int_{\CC^-} \frac{\rd \m}{2 \p \i}
             \frac{\chi (\m + 2\i, \x^-)}{\Af (\m + 2\i)} K(\la - \m) \epc
\end{align}
\end{subequations}
where $\la \in \CC^+$ in the first equation and $\la \in \CC^-$ in the
second equation.

After taking the Trotter limit $N \rightarrow \infty$ the driving terms
in the integral equations above are singular at $\x = 0$. The singularity
can be removed by taking appropriate linear combinations. We define
\begin{equation} \label{lincom}
     \begin{pmatrix}
        G^\pm (\la, \x) \\ S^\pm (\la, \x)
     \end{pmatrix} =
     \begin{pmatrix}
        \Af (\x^-)/\Bf (\x^-) & \Af (\x^+)/\Bf (\x^-) \\
        - 1/\Af (\x^+) & 1/\Af (\x^-)
     \end{pmatrix}
     \begin{pmatrix}
          \chi(\la \mp 2\i, \x^+) \\
          \chi(\la \mp 2\i, \x^-)
     \end{pmatrix} \epp
\end{equation}
Then, using also (\ref{auxid1})-(\ref{auxid3}), we arrive at equations
(\ref{gpm}), (\ref{spm}) of the main body of the text.

In order to express the determinant under the integral in (\ref{d2wide})
in terms of the functions $G^\pm$, $S^\pm$ we use the matrix $\Th^{(p)}$
defined in (\ref{matrixtheta}). Because the determinant of the matrix in
(\ref{lincom}) is $2/\Bf (\x^-)$, we obtain
\begin{equation}
     \det \Th_{j,k}^{(p)} = \biggl[ \prod_{j=1}^m \frac{2\i}{\Bf(\x_j - \i)}
                            \biggr] \det \chi_{jk}^{(p)} \epp
\end{equation}
Inserting this into (\ref{d2wide}) we arrive at the multiple integral
representation (\ref{d2}).

}

\bibliographystyle{ourbook}
\bibliography{hub}

\end{document}